\documentclass[sigconf,nonacm]{acmart}

\usepackage{common}
\usepackage{amsmath}
\usepackage{tabularx}
\usepackage{booktabs,caption}
\usepackage{longtable}
\usepackage{multirow}
\usepackage{makecell}
\usepackage{pifont}
\usepackage{subcaption}
\usepackage{xspace}
\usepackage{float}
\restylefloat{table}

\usepackage{cleveref}
\crefname{hyp}{hypothesis}{hypotheses}
\Crefname{hyp}{Hypothesis}{Hypotheses}

\newcommand{\cmark}{\ding{51}}
\newcommand{\xmark}{\ding{55}}

\AtBeginDocument{%
  \providecommand\BibTeX{{%
    \normalfont B\kern-0.5em{\scshape i\kern-0.25em b}\kern-0.8em\TeX}}}

\newcommand{\UOptimal}{\ensuremath{U_{\text{optimal}}}}
\newcommand{\RSlope}[1]{R_{\text{slope}#1}}

\newcommand{\HNull}{\textbf{H\textsubscript{0}}\xspace}
\newcommand{\HNullB}{\textbf{H'\textsubscript{0}}\xspace}

\acmYear{2020}\copyrightyear{2020}
\setcopyright{acmlicensed}
\acmConference[AFT '20]{2nd ACM Conference on Advances in Financial Technologies}{October 21--23, 2020}{New York, NY, USA}
\acmBooktitle{2nd ACM Conference on Advances in Financial Technologies (AFT '20), October 21--23, 2020, New York, NY, USA}
\acmPrice{15.00}
\acmDOI{10.1145/3419614.3423254}
\acmISBN{978-1-4503-8139-0/20/10}

\begin{document}

\title[DeFi Protocols for Loanable Funds: Interest Rates, Liquidity and Market Efficiency]{DeFi Protocols for Loanable Funds: \protect\\ Interest Rates, Liquidity and Market Efficiency}


\author{Lewis Gudgeon}
\affiliation{
    \department{Department of Computing}
    \institution{Imperial College London}
}

\author{Sam Werner}
\affiliation{
    \department{Department of Computing}
    \institution{Imperial College London}
}

\author{Daniel Perez}
\affiliation{
    \department{Department of Computing}
    \institution{Imperial College London}
}

\author{William J. Knottenbelt}
\affiliation{
    \department{Department of Computing}
    \institution{Imperial College London}
}

\renewcommand{\shortauthors}{Gudgeon et al.}

\begin{abstract}

We coin the term \textit{Protocols for Loanable Funds (PLFs)} to refer to protocols which establish distributed ledger-based markets for loanable funds.
PLFs are emerging as one of the main applications within Decentralized Finance (DeFi), and use smart contract code to facilitate the intermediation of loanable funds.
In doing so, these protocols allow agents to borrow and save programmatically.
Within these protocols, interest rate mechanisms seek to equilibrate the supply and demand for funds.
In this paper, we review the methodologies used to set interest rates on three prominent DeFi PLFs, namely Compound, Aave and dYdX.
We provide an empirical examination of how these interest rate rules have behaved since their inception in response to differing degrees of liquidity.
We then investigate the market efficiency and inter-connectedness between multiple protocols, examining first whether Uncovered Interest Parity holds within a particular protocol and second whether the interest rates for a particular token market show dependence across protocols, developing a Vector Error Correction Model for the dynamics.
\end{abstract}


\maketitle

\vspace{-2mm}
\section{Introduction}
\label{sec:introduction}

A recent development within financial architecture based on decentralized ledgers, or DeFi for short, is the emergence of protocols which facilitate programmatic borrowing and saving.
Such protocols represent a significant advancement for DeFi due to the importance of these operations to an economy.
Markets for loanable funds, a matching market for savers and would-be borrowers, in principle enable agents to engage in intertemporal consumption smoothing, whereby agents choose their present and future consumption to maximize their overall welfare~\cite{fisher1930theory}.
That is, access to loans enables a borrower to consume more today than their income would permit, paying back the loan when their income is higher.
On the other hand, savers, for whom income is higher than their present consumption, are able to deposit their funds and earn interest on them~\cite{robertson1934industrial,ohlin1937some}.

Here, we term protocols that intermediate funds between users as Protocols for Loanable Funds (PLFs).
In doing so, we note such protocols are \textit{not} directly acting as a fully-fledged replacement for banks, not least because traditional banks are not intermediaries of loanable funds: rather, they provide financing through money creation~\cite{Jakab2015} (see Section~\ref{sec:protocols}).
Further, at present PLFs only offer \emph{secured} lending, where agents can only borrow an amount provided they can front at least this amount as collateral.
This reflects the trustless setting within which PLFs operate: absent the typical repurcussions of reneging on debt commitments in traditional finance since in DeFi agents could simply default on their loans without recourse.\footnote{The enforcement of strong-identities, a mapping of on-chain to real world identities, would plausibly alter this tradeoff.}
Therefore, at present the extent to which PLFs facilitate `true' borrowing---where an agent gets into a position of net debt---is limited. 

In PLFs, interest rates reflect the prevailing \textit{price} of funds resulting from supply and demand.
The mechanism used to set these rates is therefore a crucial aspect of protocol design: it provides the pre-conditions under which the process of \textit{tatonnement}---or reaching the equilibrium---occurs~\cite{walker1987walras}.
In traditional finance, interest rates are primarily set by central banks---via a base rate---and function as a key lever in the management of credit in economies~\cite{boebaserate,federalfundsrate}. 
The lowering of the base rate makes it relatively cheaper to borrow, while discouraging saving.
In the context of PLFs, the interest rate setting mechanism is decided upon at the protocol level, commonly via a governance process.

In this paper, we seek to gain insights into how currently-deployed PLFs operate, setting out the interest rate models they employ. 
Moreover we seek to characterize the periods of illiquidity---roughly, where most of the funds within a PLF are loaned out and unavailable for withdrawal by their depositors---that these protocols have experienced. 
We then seek to understand how efficient these protocols are at present, investigating whether the no-arbitrage condition of Uncovered Interest Parity (UIP) holds \textit{within} a particular protocol. 
The efficiency of the markets serves to provide indication of the level of financial maturity, as well as the responsiveness of agents to economic incentives.
Finally, we look at the interrelation of interest rate markets across protocols, developing a Vector Error Correction Model (VECM) for the dynamics between Compound~\cite{web:compoundfinance}, dYdX~\cite{web:dydx} and Aave~\cite{web:aave} in the markets for the stablecoins DAI and USDC. 

\subsection*{Contributions}
This paper makes the following contributions:
\begin{itemize}
    \item We provide a taxonomy of the interest rate models currently employed by PLFs, resulting in three categories: linear, non-linear and kinked rates.
    \item We collect and analyze data on interest rates, utilization and the total funds borrowed and supplied on three of the largest PLFs.
    We have made the dataset publicly available.
    \item We present the first liquidity study of the markets for DAI, ETH and USDC across these PLFs, finding that periods of illiquidity are common, often shared between protocols and that liquidity reserves can be very unbalanced, with in some cases as few as \emph{three} accounts controlling c.\ 50\% of the total liquidity. 
    We also find that realized interest rates tend to cluster around the kink of a kinked interest rate model.
    \item Investigating the largest PLF, Compound, we find that the no arbitrage condition of Uncovered Interest Parity typically does \emph{not} hold, suggesting that markets associated with these protocols may be relatively inefficient and agents may not be optimally reacting to interest rate incentives.
    \item We examine the market dependence between PLFs and find that the borrowing interest rates exhibit some interdependence, with Compound appearing to influence borrowing rates on other, smaller PLFs.
\end{itemize}

The remainder of this paper is organized as follows.
Section~\ref{sec:background} presents relevant background material, while Section~\ref{sec:protocols} outlines the general design of PLFs.
Section~\ref{sec:models-tradeoffs} presents a taxonomy of different interest rate models.
Sections~\ref{sec:empirical}, ~\ref{sec:market-efficiency} and ~\ref{sec:market-dependence} provide an analysis on market liquidity, efficiency and dependence, respectively.
Section~\ref{sec:related_work} discusses related work, before Section~\ref{sec:conclusion} concludes.

\section{Background}
\label{sec:background}

\subsection{Ethereum}
The Ethereum~\cite{wood2014ethereum} blockchain allows its users to run \textit{smart contracts}, programs designed to run on its distributed infrastructure.
Smart contracts and the interactions between them are fundamental building blocks of DeFi.
They are almost feature-equivalent to programs written in any Turing-complete language but have a few particularities.
For instance, smart contracts must be strictly deterministic.
For this reason, they can only communicate with the outside world through transactions executed on the Ethereum blockchain.
On the other hand, smart contracts can easily interact with other smart contracts, allowing complex interactions between different parties as long as these interactions happen directly on chain.
Another particularity of Ethereum smart contracts is their atomicity: they can only be executed within a transaction. If an error happens during the execution, the transaction is reverted.
In such an event, any change of state that occurred in this contract or any other interaction with other contracts will be reverted and no change of state will happen.

\subsection{DeFi}
DeFi refers to a financial system which relies for its security and integrity on distributed ledger technology.
Applications of such technology include lending, decentralized exchange, derivatives and payments.
At the time of writing on 9 June 2020, DeFi has a total value locked of over 1bn USD, with most applications deployed on the Ethereum blockchain~\cite{defipulse}.
Unlike regular finance where the identity of all participants is known and correct behavior can be enforced via regulation, DeFi actors are pseudonymous and DeFi systems need other means to prevent users from misbehaving.
In the absence of traditional credit-rating mechanisms, the system rules are typically ``enforced'' by incentivizing actors to behave according to the rules of the system~\cite{gudgeon2020decentralized}.

\subsection{DeFi lending markets}
PLFs intermediate markets for loanable funds, with suppliers of funds earning interest.
As mentioned above, protocols need to protect against borrowers defaulting on their debt obligations.
Where loans need to be valid for more than a single transaction, this protection is currently achieved by requiring borrowers to \textit{over-collateralize} their loans, allowing the lender to redeem the pledged collateral should a borrower default on a position\footnote{Therefore loans of this type on DeFi lending protocols are instances of secured loans, where an agent can only borrow against collateral they already own; they cannot enter into `net debt'. We address this further in Section~\ref{sec:protocols}.}.
Where the loan needs to be valid only for a single transaction, \textit{flash loans} enable agents to borrow without collateral, whereby the loaned amount is protected by the atomicity afforded by smart contracts: if the loan is not repaid with interest, the whole transaction is reversed~\cite{web:aave}.

In the context of lending protocols, a borrower defaults on a loan when the value of the locked collateral drops below some fixed liquidation threshold.
The liquidation thresholds vary between asset markets across different protocols.
In an event of default, the lending protocol seizes and liquidates the locked collateral at a discount to cover the underlying debt.
Additionally, a penalty fee is charged against the debt, prior to paying out the remaining collateral to the borrower.   

\subsection{Stablecoins}
In order for a cryptoasset to be a viable medium of exchange and store of value, price stability needs to be guaranteed. 
\textit{Stablecoins} are cryptoassets which possess a price stabilization mechanism to maintain some target peg.
Here we briefly outline two of the most widely used stabilization mechanisms~\cite{moin2019sok}:

\begin{description}
    \item[Fiat-collateralized.]
    Each unit of stablecoin is pegged to some fixed amount of fiat currency (typically USD).
    This is generally realized via a network of banks maintaining the fiat collateral and is therefore not decentralized. 
    Stablecoins such as USDT~\cite{whitepaper:tether} and USDC~\cite{web:usdc} belong in this category.
    \item[Cryptoasset-collateralized.]
    Each unit of stablecoin is backed by an amount of some other cryptoasset.
    A stabilization mechanism is needed to protect against the volatility of the collateral.
    Perhaps the most prominent of such stablecoins is DAI~\cite{whitepaper:maker}.
    In order to borrow newly minted units of DAI, where one DAI is pegged to 1 USD, a user has to pledge an over-collateralized amount of cryptocurrency (e.g. ETH), which becomes locked up in a smart contract.
    In case the price of DAI deviates from its peg, arbitrageurs are incentivized to buy or sell DAI should the price drop below or rise above 1 USD, respectively.
    A borrower of DAI has to ensure to keep the associated collateralization ratio above some liquidation threshold, as otherwise the borrow position will be liquidated at a discount and a penalty fee will be charged against the debt.
\end{description}

\section{Protocols for loanable funds}
\label{sec:protocols}

\subsection{Comparison to traditional lending}
PLFs facilitate the matching of would-be borrowers and lenders, with the interest rate set programmatically.
Importantly, unlike peer-to-peer lending, funds are pooled, such that a lender may lend to a number of borrowers and vice versa.
In so doing, an open lending protocol provides a market for loanable funds, where the role that an intermediary would play in traditional finance has been replaced by a set of smart contracts.

It should be stated that by creating markets for loanable funds---as protocols for loanable funds---such protocols are not functionally equivalent to banks. 
The construal of banks as primarily intermediaries of loanable funds (ILFs), as in some economic theory, has been debunked (see e.g.\ ~\cite{Jakab2015}).
Rather than accepting deposits of pre-existing funds from savers and then lending these funds out to borrowers, banks primarily provide financing through \textit{money creation}, creating new money at the point of making a loan and constrained by their profitability and solvency requirements~\cite{Jakab2015}.
Therefore since banks are not primarily ILFs, PLFs are not functional replacements. 

\subsection{Use cases}
The introduction of PLFs significantly extends the existing trading capabilities in DeFi, offering several use cases for DeFi actors.
Predominantly, PLFs empower decentralized \textit{margin trading} by facilitating \textit{short sells} and \textit{leveraged longs}.
Margin is defined as the collateral that an agent gives to a counterparty in order to cover the credit risk that the agent poses for the counterparty.
In a short sell, a trader sells the borrowed funds, seeking to make a profit by repurchasing the borrow position at a lower price.
Similarly, in a leveraged long a trader buys some other asset with the borrowed funds and profits in case the purchased asset appreciates in value.
As a consequence of margin trading, suppliers of loanable funds are able to earn interest.

A further use case of PLFs lies in borrowers being able to leverage their funds as collateral, while maintaining the right to repurchase the collateralized token, thereby not giving up direct ownership. 
\subsection{Design space dimensions}

\subsubsection{Interest rate model}
Suppliers of loanable funds receive interest over time, while borrowers have to pay interest.
A key differentiating factor across lending protocols is the chosen interest rate model, which is generally some linear or non-linear function of the available liquidity in a market.
As loans on protocols for loanable funds have unlimited maturities, variable interest rates may fluctuate from the opening of a borrow position.
By using variable rate models, lending protocols are able to dynamically adjust the interest rate depending on the ratio of funds borrowed to supplied, which can prove particularly useful during periods of low liquidity by incentivizing borrowers to repay their loans.

\subsubsection{Reserve factor}
Additionally, lending protocols employ a \textit{reserve factor}, specifying the amount of a borrower's accrued interest to be deducted and set aside for periods of illiquidity. 
Hence, the interest earned by lenders is a function of the interest paid by borrowers less the reserve factor.

\subsubsection{Interest disbursement mechanism}
Interest is typically accrued per second and paid out on a per block basis.
Since the repeated payment to lenders of the accrued interest (denoted in the supplied token) would incur undesired transaction costs, accrued interest is often paid out through the use of \textit{interest-bearing derivative tokens}, which are ERC-20 tokens that are minted upon the deposit of funds and burned when redeemed.
Each market has such an associated derivative token, which appreciates with respect to the underlying asset at the same rate as interest is compounded, thereby accruing interest for the token holder. 
Even though loans are made with indefinite maturity, a loan is liquidated should the value of the borrowed asset's underlying collateral fall below a fixed liquidation threshold.
In the case of an undercollateralized borrow position, so-called liquidators can purchase the collateral at a discount and a penalty fee is imposed upon the borrower.

\subsubsection{Governance mechanism}
A critical component of lending protocols is decentralized governance.
Lending protocols tend to achieve decentralized governance through the use of ERC-20 governance tokens specific to the lending protocol, whereby token holders' votes are weighted proportionally to their stake.
Token holders are thereby empowered to propose new features and changes to the existing protocol.

\begin{table*}[t]
  \centering
  \begin{tabular}{lcccccc}
    \thead[l]{Protocol} & \thead[c]{Interest Rate \\Model} & \thead[c]{Stable \\Interest Rate} & \thead[c]{Variable \\Interest Rate} & \thead[c]{Governance \\Token} & \thead[l]{Interest-bearing\\ Derivative Token} & \thead[c]{Additional \\Functionalities}\\
    \toprule    
    Compound & Kinked & \xmark & \cmark & \cmark & \cmark & --\\
    Aave & Kinked & \cmark & \cmark & \cmark & \cmark & Swap rates, flash loans\\
    dYdX & Non-linear & \xmark & \cmark & \xmark & \xmark & Decentralized exchange, flash loans\\
    \bottomrule
  \end{tabular}
  \caption{Comparison of different protocols for loanable funds.}
  \label{tab:protocols}
\end{table*}

\section{Interest rate models}
\label{sec:models-tradeoffs}

In this section, we outline the main classes of interest rate models employed by PLFs.
The interest rate model used can differ both across PLFs and by market within a particular PLF. 
We also describe an approach that has been taken to enable these variable rate models to offer more interest rate stability.  

\paragraph{Definitions.}
For a market $m$, total loans $L$ and gross deposits $A$, we define the utilization of deposited funds in that market as
\begin{equation}
\label{eqn:compound-finance}
    U_m = \frac{L}{A}
\end{equation}
The Interest Rate Index $I$ for block $k$ is calculated each time an interest rate changes, i.e. as users mint, redeem, borrow, repay or liquidate assets.
It is given by:
\begin{equation}
\label{eqn:compound-interest-rate-index}
I_{k,m} = I_{k-1,m} (1+rt)
\end{equation}
where $r$ denotes the per block interest rate and $t$ denotes the difference in block height.
Therefore debt $D$ in a market is given by
\begin{equation}
\label{eqn:compound-debt}
D_{k,m} = D_{k-1,m} (1+rt)
\end{equation}
where a portion of the interest is kept as a reserve ($\Pi$), set by reserve factor $\lambda$:
\begin{equation}
\label{eqn:compound-reserves}
\Pi_m = \Pi_{k-1, m} + D_{k-1,m} (rt\lambda)
\end{equation}

We now turn to the classification of the extant interest rates into three main models. 

\subsection{Model one: linear rates}
\label{sec:linear-rates}



The first model we present is one in which interest rates are set as a linear function of utilization. 
With a linear interest rate model, interest rates are determined algorithmically as the equilibrium value in a loanable market $m$, where the borrowing interest rates $i_b$ are given by:

\begin{equation}
\label{eqn:compound-borrow-rate}
i_{b,m} = \alpha + \beta U_m
\end{equation}

where $\alpha$ is some constant and $\beta$ a slope coefficient on the responsiveness of the borrowing interest rate to the utilization rate. 
Saving interest rates $i_s$ are given by:

\begin{equation}
\label{eqn:compound-saving-rate}
i_{s,m} = (\alpha + \beta U_m)U_m
\end{equation}

where in essence the interest rate $i_{b,m}$ is scaled by the utilization to arrive at an interest rate for saving that is lower than that of the rate paid by borrowers. 
This serves to ensure that the interest rate spread ($i_{b,m} - i_{s,m}$) is positive.
Some portion of this spread can be kept for reserves. 


\subsection{Model two: non-linear rates}
\label{sec:non-linear-rates}



Interest rates may also be set non-linearly, and here we present the non-linear model employed by dYdX~\cite{web:dydx-sourcecode}.
For a loanable funds market $m$, the borrowing interest rates $i_b$ follow a non-linear model and are computed as:

\begin{equation}
    i_{b,m} = (\alpha \cdot U_m) + (\beta \cdot U_m^{32}) + (\gamma \cdot U_m^{64}).
\end{equation}

The saving interest rates $i_s$ with reserve factor $\lambda$ are given by:

\begin{equation}
    i_{s,m} = (1-\lambda) \cdot i_{b,m} \cdot U_m
\end{equation}

In comparison to the linear rate model, a non-linear model allows for the interest rate to increase at an increasing rate as the protocol becomes more heavily utilized, creating an non-linearly increasing incentive for suppliers to supply to the protocol and for borrowers to repay their borrows. 


\subsection{Model three: kinked rates}
\label{sec:kinked-rates}

In the final interest rate model, interest rates exhibit some form of kink: they sharply change at some defined threshold.
Such interest rates are employed by a number of protocols, including~\cite{web:compoundfinance, web:compound-sourcecode,web:aave-interest-documentation, web:aave-sourcecode}.

Mathematically, kinked interest rates can be characterized as follows. 

\begin{equation}
    i_b =
    \begin{cases}
      \alpha + \beta U & \text{if}~U \leq U^{*}\\
      \alpha + \beta U^{*} + \gamma (U-U^{*}) & \text{if}~U > U^{*}
    \end{cases}
    \label{eq:compound-borrow}
  \end{equation}
  
where $\alpha$ denotes a per-block base rate, $\beta$ denotes a per-block multiplier, $U$ denotes the utilization ratio (with $U^{*}$ denoting the optimal utilization ratio) and $\gamma$ denotes a `jump' multiplier.

In the case of Compound, the associated saving rates are given by Equation~(\ref{eq:compound-save}).

\begin{equation}
    i_s = U (i_b(1-\lambda))\\
    \label{eq:compound-save}
\end{equation}

where $\lambda$ is a reserve factor. 

Such models share the property of sharply changing the incentives for borrowers and savers beyond some utilization threshold, as with the non-linear model. 
However, they also introduce a point of sharp change in the interest rate, beyond which the interest rates starts to sharply rise, in contrast to non-linear models with no kink. 
Therefore it might be expected that this kink would become a Schelling point\footnote{Informally, a solution of a coordination game that agents tend to arrive at in the absence of communication, such as two strangers who wish to meet but cannot communicate deciding to meet at noon at the Grand Central Terminal in New York City, since this somehow seems a \emph{natural} choice\cite{schelling1958strategy}.} of convergence among agents~\cite{schelling1958strategy}.







\subsection{Making rates stable}
Some platforms, such as Aave, allow the borrower to \emph{choose} between a variable and a stable interest rate.
However, it is important to note that the ``stable'' interest rate is not \textit{entirely} stable, as it can be revised in the event that it significantly deviates from the market average.
Examining Aave's implementation in detail, we first present their instantiation of a kinked interest rate model before showing how the stable rate is derived\footnote{These formulae are an adapted version of those that appear in the Aave whitepaper~\cite{web:aave}}. 

The variable interest rate is based on several parameters defined by the system.
Given the utilization rate $U$ of a particular asset, the parameter $\UOptimal$ is the optimal utilization. 
In practice, this value was set to $0.8$ and has been updated to $0.9$ in May 2020~\cite{aave-rates-update}.
Two interest rate slopes, parameters of the system, are used to compute the variable interest rate: $\RSlope{1}$ is used when $U < \UOptimal$ and $\RSlope{2}$ when $U \geq \UOptimal$.
Finally, given a base variable borrow rate $i_{b,m,v_{0}}$, the variable borrow interest rate $i_b$ for market $m$ is computed as follows:

\begin{equation}
  i_{b,m,v} =
  \begin{cases}
    i_{b,m,v_{0}} + \frac{U}{\UOptimal}\cdot \RSlope{1} & \text{if}~U < \UOptimal\\
    i_{b,m,v_{0}} + \RSlope{1} + \frac{U - \UOptimal}{1 - \UOptimal}\cdot \RSlope{2} & \text{if}~U \geq \UOptimal
  \end{cases}
\end{equation}

To compute the stable rate, Aave computes the lending protocol-wide market rate $m_r$ as the arithmetic mean of the total borrowed funds weighted by the borrow rate $i_{b,m}$ for given platform $p$ as follows:

\begin{equation}
  m_r = \frac{\sum_{p=1}^{n} i_{b,m,p} \cdot B_{m,p}}{\sum_{p=1}^{n} B_{m,p}}
\end{equation}

where $B_{m,p}$ denotes the total amount of borrowed funds for market $m$ on lending protocol $p$.
Hence, using the $m_r$ as the base rate, the stable borrowing rate $i_{b,s}$ for a market $m$ is given by:

\begin{equation}
    i_{b,m,s} = 
    \begin{cases}
        m_r + \frac{U}{\UOptimal}\cdot \RSlope{1} & \text{if}~U < \UOptimal\\
        m_r + \RSlope{1} + \frac{U - \UOptimal}{1 - \UOptimal}\cdot \RSlope{2} & \text{if}~U \geq \UOptimal 
    \end{cases}
\end{equation}

In case the stable rate deviates too much from the market rate, it will be revised. 
The stable borrow rate $i_{b,m,s}$ for user $z$ is revised upwards to the most recent stable borrow rate for the respective market when

\begin{equation}
\label{aave:down}
    i_{b,m,s,z} < \frac{B_{m,v}\cdot i_{b,m,v} + B_{m,s}\cdot i_{b,m,s}}{B_{m,v} + B_{m,s}}
\end{equation}

If \eqref{aave:down} holds, a borrower of funds would be able to earn interest from a borrow position.
On the contrary, should the stable rate of a borrow position exceed the latest stable rate it would be adjusted downwards should

\begin{equation}
    i_{b,m,s,z} > i_{b,m,s}\cdot(1 + \Delta i_{b,m,s,t})
\end{equation}

where $\Delta i_{b,m,s,t}$ denotes the change in the stable rate for a specified adjustment window $t$.
Note that unlike for variable interest rate denominated loans, stable rate loans have a definite maturity. 

\subsection{Summary}

We have reviewed the three main interest rate models for variable interest rates, and explained a mechanism which seeks to bring stability to these rates.
An emergent key feature of these models is the incentive they provide to borrowers and savers at times of high utilization. 
In the next section, this behavior at high utilization becomes a central object of concern.

\section{Market liquidity}
\label{sec:empirical}
In this section we provide an analysis of liquidity and interest rates for loanable funds markets on Compound, dYdX and Aave.

\subsection{Liquidity and illiquidity across PLFs}
\label{sec:liqudity-illiquidity}
The total amount of locked loanable funds for the largest markets across Compound, Aave and dYdX are given in Table \ref{tab:total-liquidity}.

\begin{table}[H]
    \centering
    \begin{tabular}{l r r r}
    \toprule
    Currency & \multicolumn{3}{c}{Total Amount Locked} \\
    &  \multicolumn{3}{c}{(median in millions of USD)}\\
             & Compound & Aave & dYdX\\
    \midrule
    (W)ETH & 76.58 & 4.80 & 19.41\\
    USDC   & 31.54 & 4.12 & 6.58\\ 
    DAI    & 24.82 & 0.95 & 4.64\\
    SAI    & 36.94 & - & -\\
    USDT   & - & 3.92 & -\\
    BAT    & 0.95 & 0.08 & -\\
    LEND   & - & 3.60 & -\\
    LINK   & - & 12.21 & -\\
    \bottomrule
    \end{tabular}
    \caption{Median of total supply of loanable funds in USD for the largest markets on Compound, Aave and dYdX, since each market's inception until 7 May 2020.}
    \label{tab:total-liquidity}
\end{table}

\begin{figure}[tbp]
    \centering
    \includegraphics[width=0.47\textwidth]{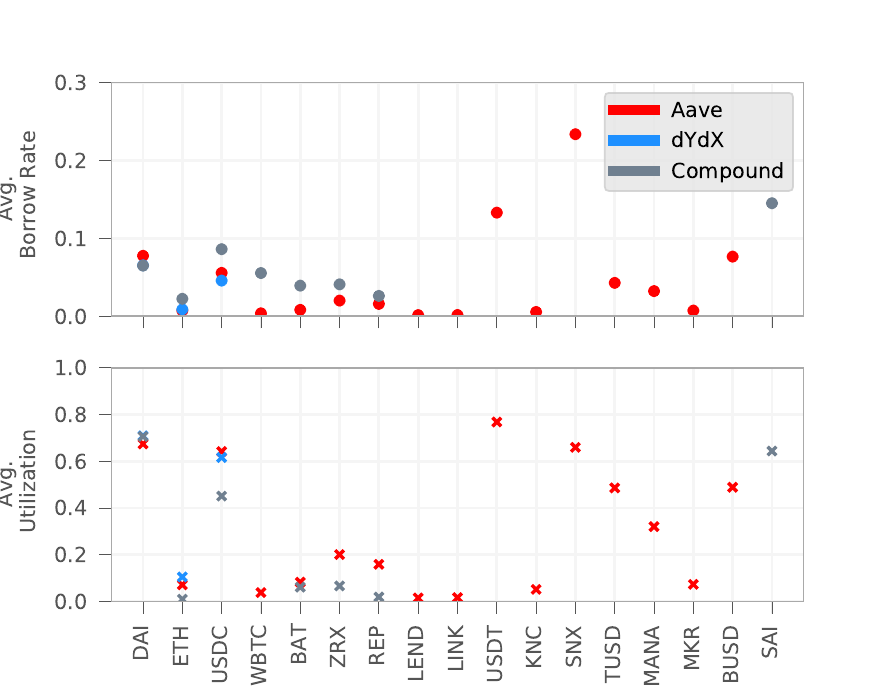}
    \caption{Average utilization and borrow interest rates for all markets on Aave, Compound and dYdX.}
    \label{fig:platforms}
\end{figure}

It can be seen that ETH, USDC and DAI account for the majority of loanable funds on all three PLFs.\footnote{As single-collateral DAI (SAI) has been replaced by multi-collateral DAI (DAI), we solely focus on the latter for this analysis.}
Hence we focus on these markets for an in-depth analysis.
From Figure~\ref{fig:platforms} it becomes apparent that these three markets are very similar in terms of their average borrow and utilization rates, particularly for DAI and ETH.

\subsubsection{Liquidity}
\begin{figure}
  \centering
  \includegraphics[width=0.47\textwidth]{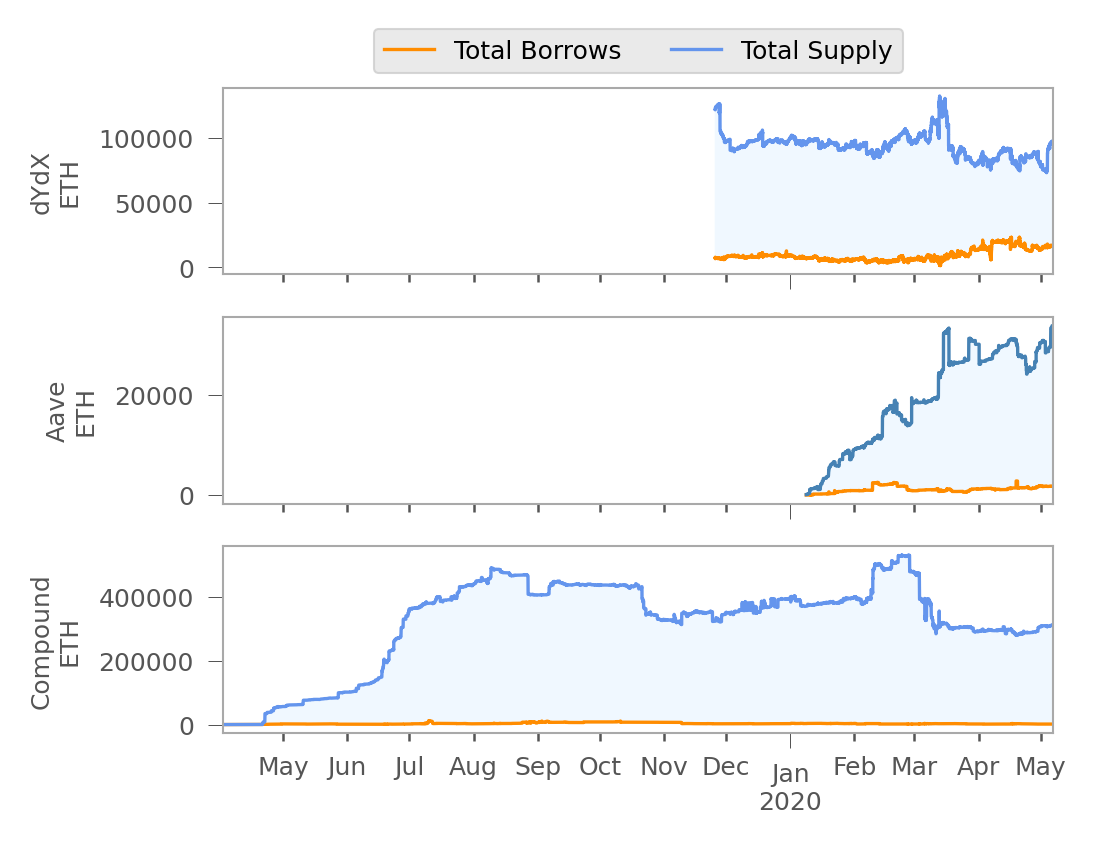}
  \caption{Total funds borrowed and supplied (i.e. liquidity) for ETH markets on dYdX, Compound and Aave.}
  \label{fig:eth-liquidity}
\end{figure}

\begin{figure}
  \centering
  \includegraphics[width=0.47\textwidth]{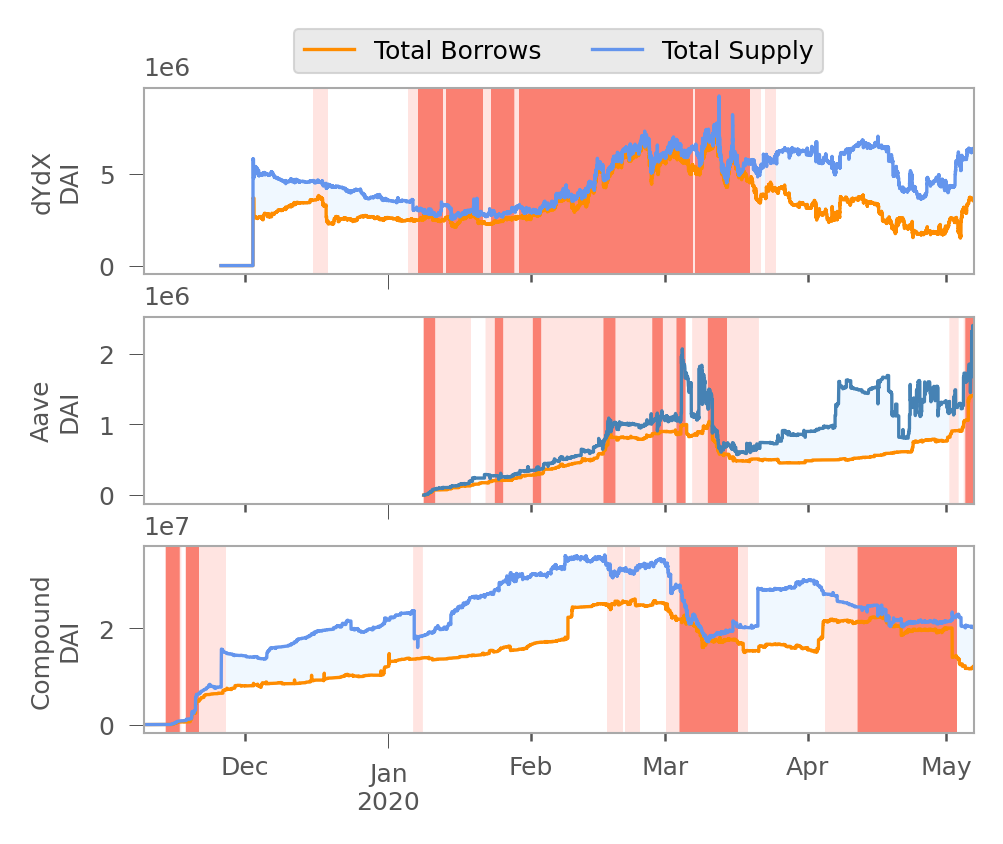}
  \caption{Total funds borrowed and supplied (i.e. liquidity) for DAI markets on dYdX, Compound and Aave. Periods where utilization was between 80\% and 90\% are highlighted in salmon, while utilization higher than 90\% is shaded in red.}
  \label{fig:dai-liquidity}
\end{figure}

\begin{figure}
  \centering
  \includegraphics[width=0.47\textwidth]{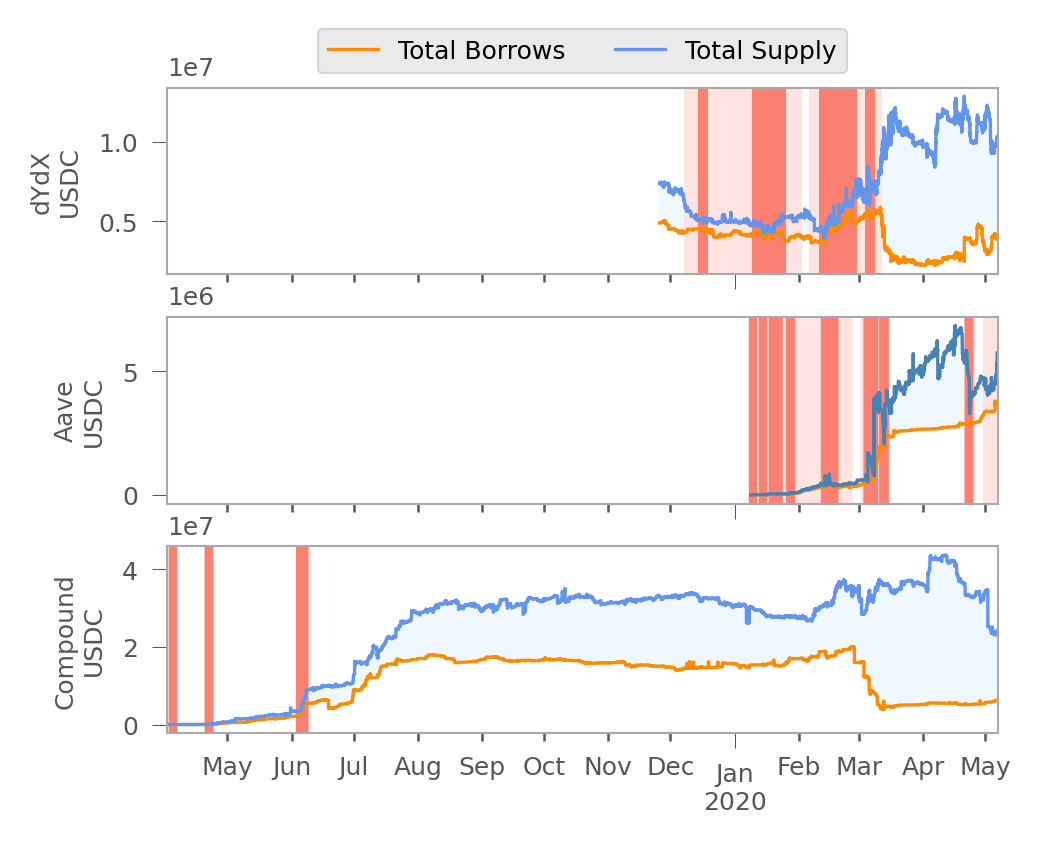}
  \caption{Total funds borrowed and supplied (i.e. liquidity) for USDC markets on dYdX, Compound and Aave. Periods where utilization was between 80\% and 90\% are highlighted in salmon, while utilization higher than 90\% is shaded in red.}
  \label{fig:usdc-liquidity}
\end{figure}

The available liquidity for loanable funds for an asset is given by the difference between the total supply and total borrows in the respective market.
High liquidity allows actors to borrow funds at lower rates, while guaranteeing suppliers of funds that funds can be withdrawn at any point in time.
On the one hand, regarding the liquidity for ETH (see Figure~\ref{fig:eth-liquidity}) all three PLFs maintain high liquidity over time, largely due to the total borrows remaining relatively stable.
On the other hand, the markets for DAI and USDC (see Figures~\ref{fig:dai-liquidity} and \ref{fig:usdc-liquidity}) frequently exhibit periods of much lower liquidity, with utilization exceeding 80\% and 90\% respectively.
Moreover, it appears that such periods of low liquidity are to some extent shared across protocols, in particular for the smaller PLFs dYdX and Aave for the period January to mid-March 2020.

On Thursday March 12, 2020---\textit{Black Thursday}~\cite{black-thursday}---the total amount of locked funds across all DeFi protocols dropped from 897.2m USD to 559.42m USD.\footnote{Source: \url{https://defipulse.com}. Accessed: 05-06-2020.}
For DAI, it can be seen how on Black Thursday even the largest PLF, Compound, was exposed to prolonged periods of low liquidity, before attracting increased liquidity again at the same time as dYdX and Aave. 
However, after mid-April, the market for DAI on Compound re-experienced low liquidity. 

\subsubsection{Illiquidity}
On PLFs agents are incentivized to provide liquidity via the employed interest rate model, as high interest rates would make borrowing more cost prohibitive in periods of low liquidity.
However, if borrowers are not incentivized to repay their loans by sufficiently high interest rates at times of full utilization, insufficient liquidity may materialize. 
In the event of such illiquidity materializing, suppliers of funds would be unable to withdraw them, being forced to hold on to and continue to earn interest through their cTokens. 

\begin{figure}
  \centering
  \includegraphics{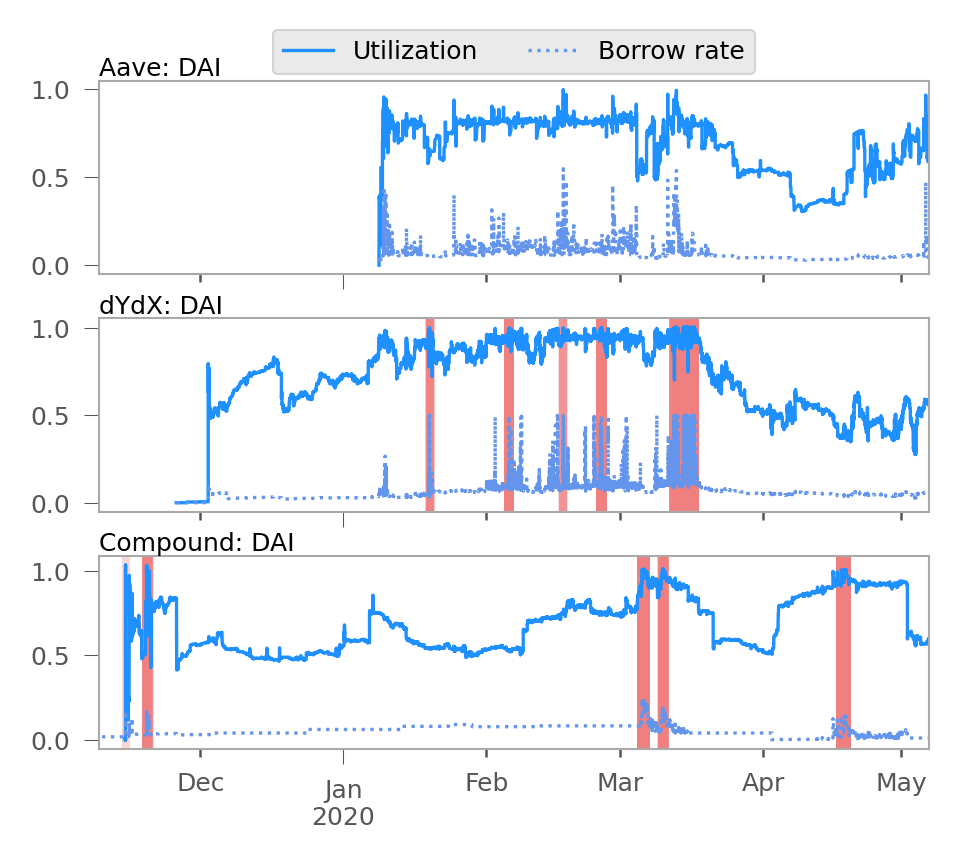}
  \caption{Utilization and borrow rates for DAI on Aave (top), dYdX (middle) and Compound (bottom). Time periods in which utilization equaled or exceeded 100\% are highlighted in red.}
  \label{fig:dai-illiquidity}
\end{figure}

Out of the three PLFs, only Aave enforces a utilization ceiling at 100\%, while Compound and dYdX permit borrows even \textit{beyond} full utilization. 
When examining the market for DAI in Figure~\ref{fig:dai-illiquidity}, it can be seen how utilization of funds has in the past been multiple times at and even above 100\% on Compound and dYdX.

It can be seen that Aave has experienced periods of near-illiquidity, while Compound and dYdX have experienced periods of full illiquidity for DAI, i.e. all supplied funds were loaned out.
When comparing the DAI borrow rates during periods of full utilization (red) in Figure~\ref{fig:dai-illiquidity}, notable differences can be made out between the different interest rate regimes.
On dYdX, the borrow rate hits the by the model imposed interest rate ceiling of 50\%, while on Compound, the rate does not exceed 25\% even at full utilization, which can be explained by the linear nature of Compound's interest rates.
Despite Aave never reaching full utilization for DAI, due to an optimal utilization target of 80\% during the measurement period, borrow rates on Aave exceed rates on Compound during periods of high utilization.
This suggests that holding on to loans during periods of illiquidity is notably cheaper on Compound than on dYdX or Aave.

\subsubsection{Fund distribution}
Periods of low liquidity have several implications for market participants. 
On one side, high utilization implies lucrative interest rates for suppliers of funds, thereby attracting new liquidity.
On the other hand, suppliers are faced with the risk of being unable to redeem their funds, for example, in the case of a `bank run'.

\begin{figure}
  \centering
  \begin{subfigure}[b]{0.47\textwidth}
     \centering
     \includegraphics[width=1\linewidth]{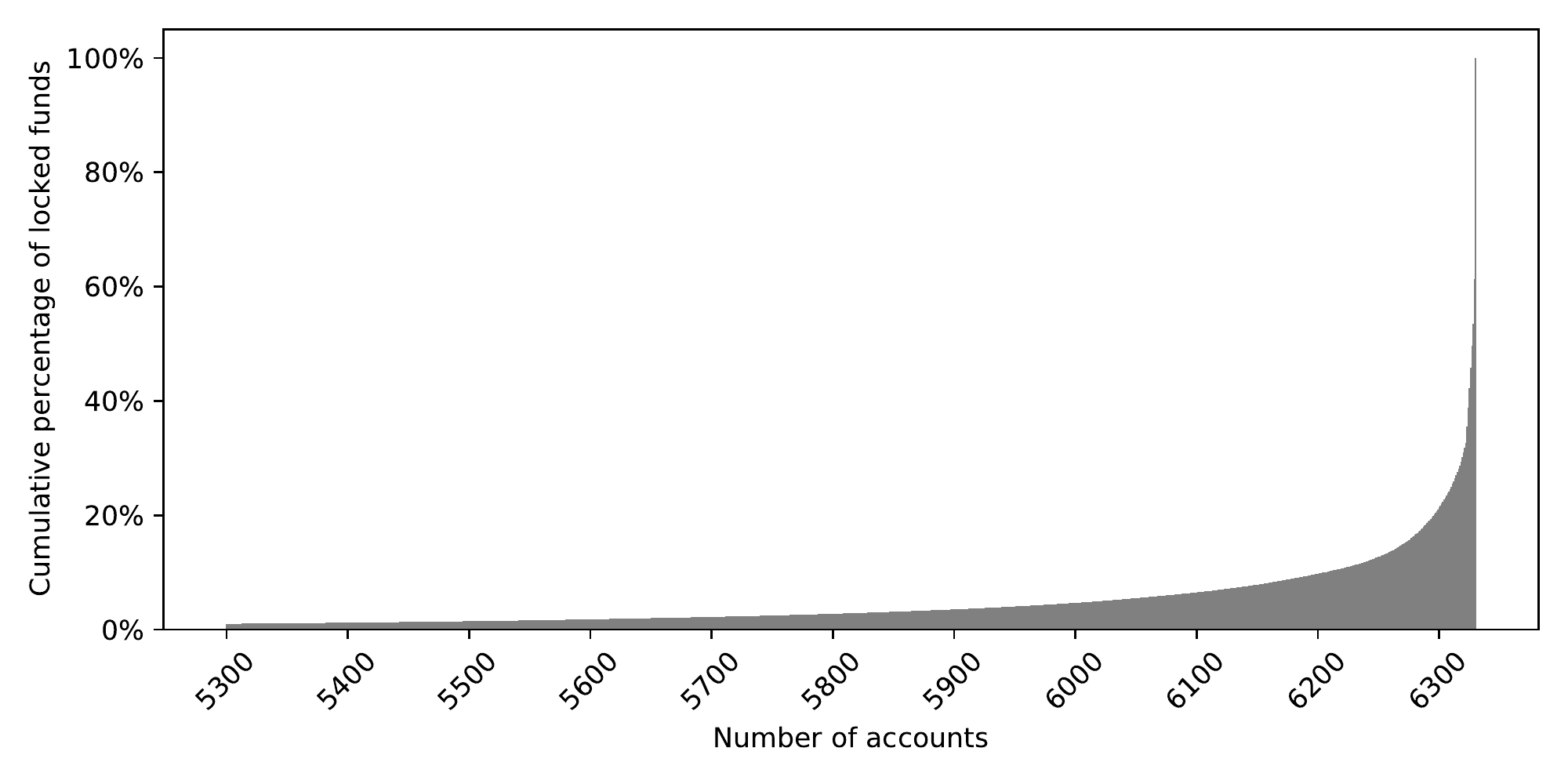}
     \caption{DAI}
     \label{fig:cdai-wealth} 
  \end{subfigure}
  
  \begin{subfigure}[b]{0.47\textwidth}
     \centering
     \includegraphics[width=1\linewidth]{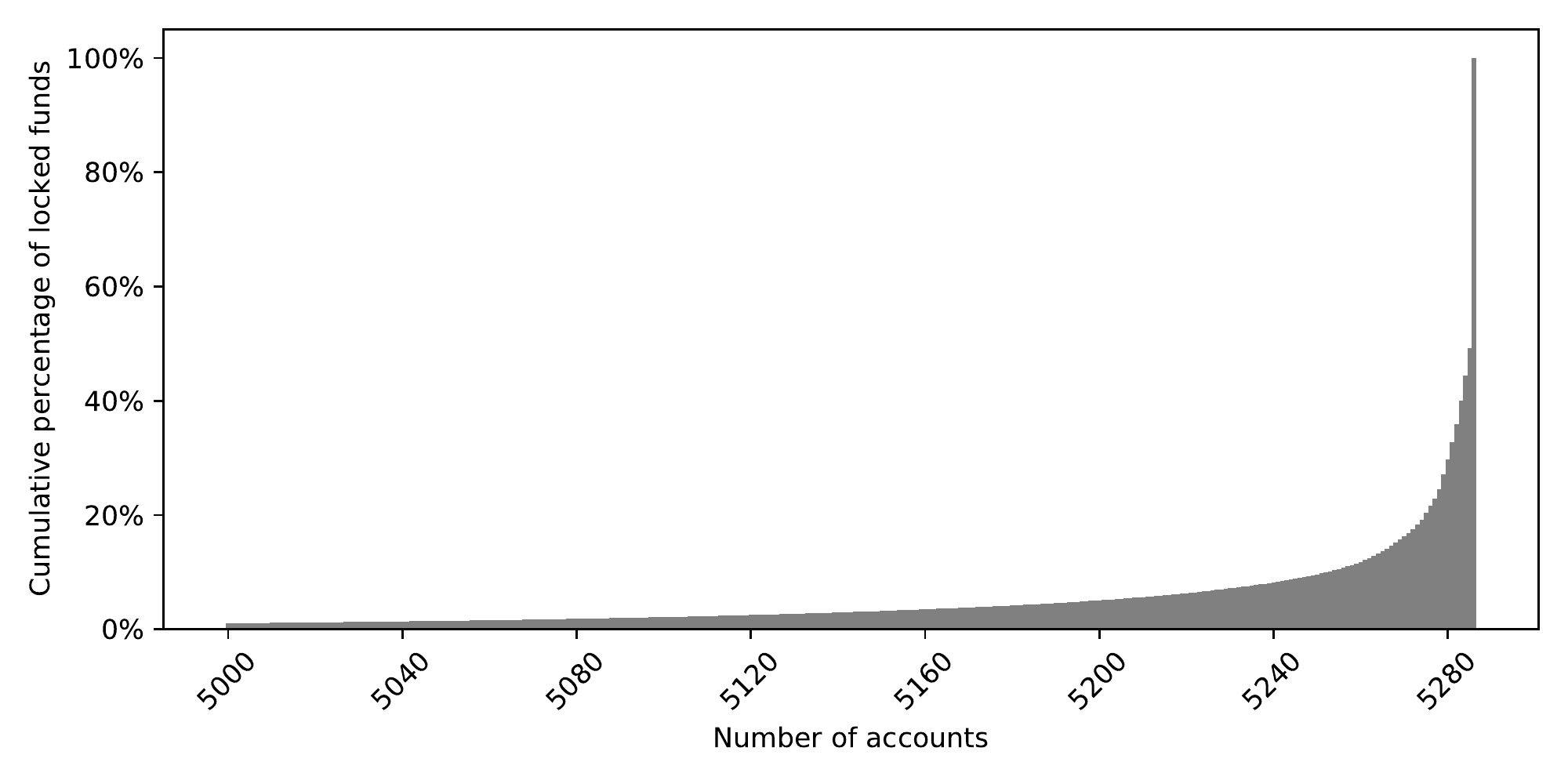}
     \caption{ETH}
     \label{fig:ceth-wealth}
  \end{subfigure}
  
  \begin{subfigure}[b]{0.47\textwidth}
    \centering
    \includegraphics[width=1\linewidth]{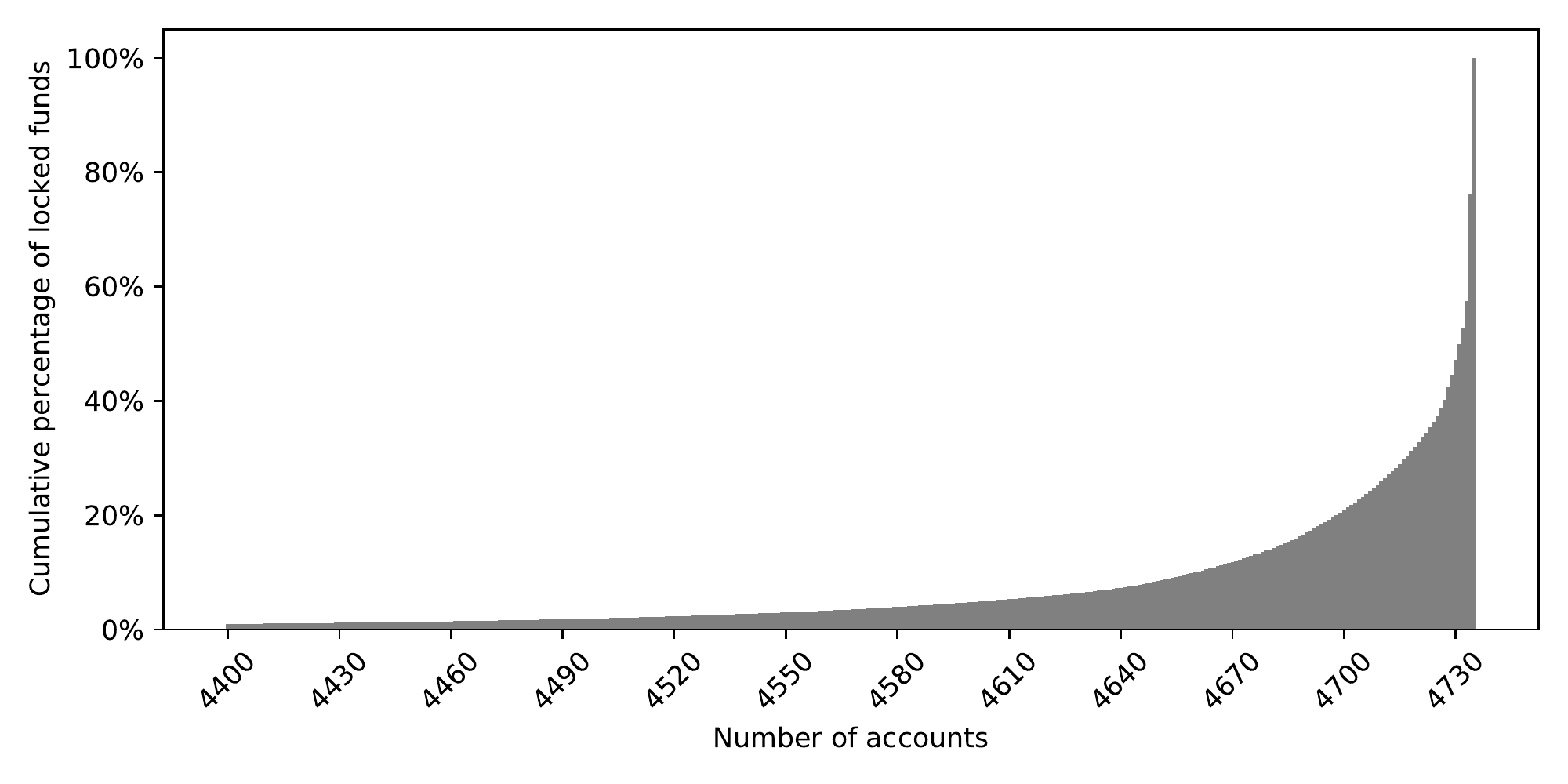}
    \caption{USDC}
    \label{fig:cusdc-wealth}
  \end{subfigure}
  
  \caption[]{Cumulative percentage of locked funds on Compound for DAI, ETH and USDC on 2020-06-04.}
  \label{fig:cumul-wealth}
  \end{figure}

In order to better assess the risk of a market becoming fully illiquid, we examine the cumulative percentage of locked funds for the number of Ethereum accounts on Compound in Figure~\ref{fig:cumul-wealth}.
Note that as a similar pattern was found for Aave and dYdX, we decided to solely focus on Compound.
The distribution of funds across accounts is very similar for DAI, ETH and USDC in that a very small set of accounts controls the majority of all supplied funds.
For instance, 50.3\% of total locked DAI is controlled by only 3 accounts.
Similarly, for ETH and USDC, the same number of accounts control 60.0\% and 47.3\%, respectively.
Hence, for all three markets, even in times of high liquidity, a small number of suppliers of funds are in a position to to drastically reduce liquidity, or possibly even cause full illiquidity.

\subsection{\textit{Case Study}: DAI on Compound}
\label{sec:behavior}

In the context of liquidity, we present a case study of interest rate behavior in the market for DAI on Compound, focusing on the period of 21 February to 21 April 2020 and its interest-bearing token cDAI. 
It could be seen in Figure~\ref{fig:dai-illiquidity} that for the aforementioned period, this market was exposed to a range of different utilization levels, experiencing periods of relatively high liquidity but also illiquidity. 
Hence, we investigate market participants' behavior---given by the interest rates that are actually observed---for different interest rate regimes during the period of interest. 

\paragraph{Interest rate models for the cDAI contract}
To illustrate kinked rates, we present the case of the DAI interest rate in Compound Finance.
The cDAI token is an example of an interest-bearing derivative token based on a linear kinked interest rate model. 
Since the 17th December 2019, the borrowing rates ($i_b$) have operated with Equation (\ref{eq:compound-borrow}).
However, the precise parameter values used by the model have been revised multiple times. We include a list of these modifications in Appendix Table \ref{tab:cdai-models}.

\paragraph{Interest rate behavior}
We consider in detail how since 17 December 2019 agents have optimized their selection of borrowing and saving amounts given an interest rate schedule.
Here we focus on a subset of three periods, namely:

\begin{itemize}
\item 21 February - 13 March 2020
\item 14 March - 5 April 2020
\item 6 April - 21 April 2020
\end{itemize}

\begin{figure}[tbp]
  \centering
  \includegraphics[width=.9\columnwidth]{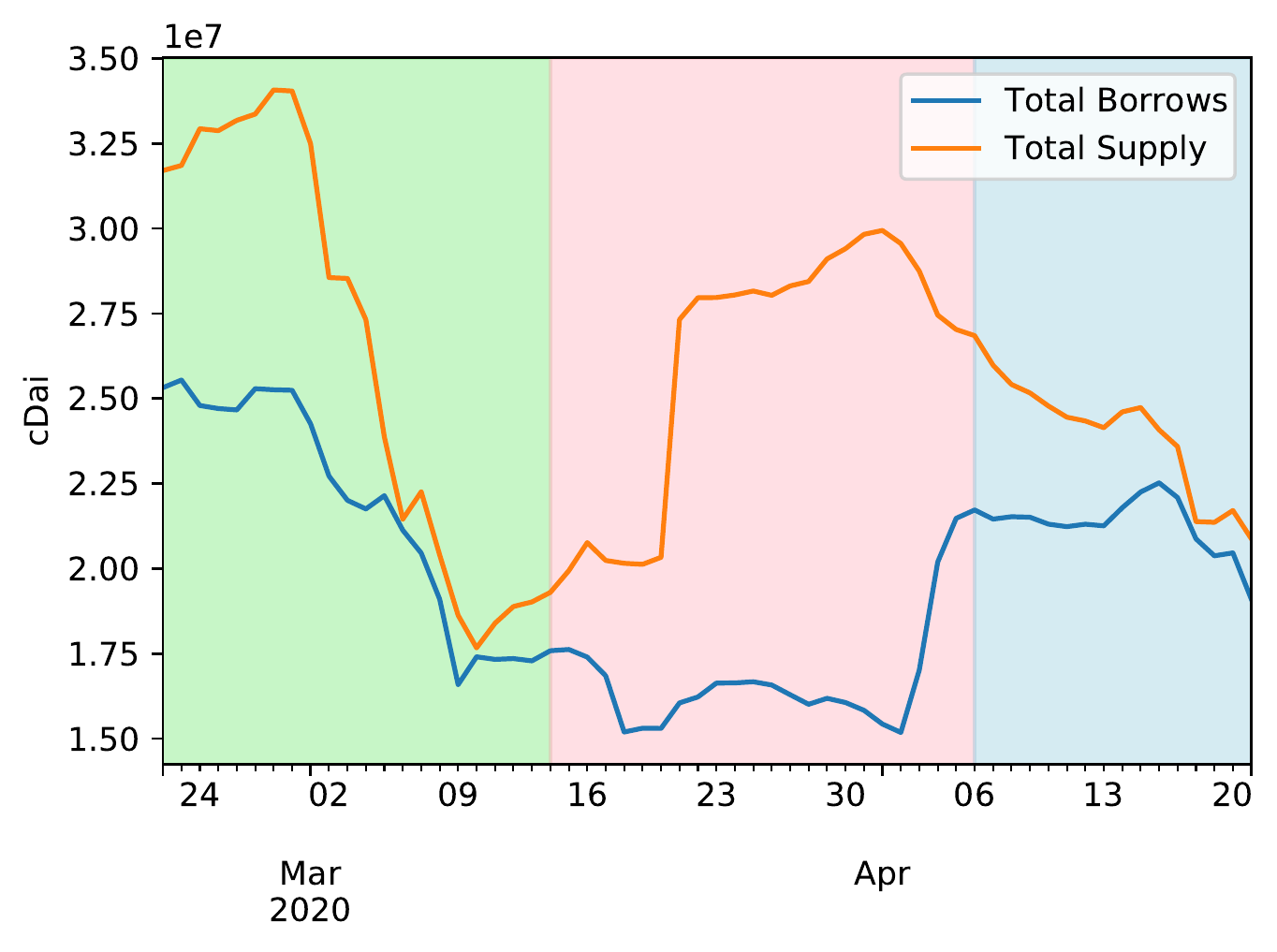}
  \caption{Three interest rate regimes in Compound.}
  \label{fig:compound-three-regimes}
\end{figure}

At the start of each of these periods, the interest rate parameters were changed to values as specified in Appendix Table~\ref{tab:cdai-models}.
Here we plot the behavior of the borrowing rates, but the behavior for the supply rates is broadly similar.

\begin{figure}
  \centering
  \begin{subfigure}[b]{0.47\textwidth}
     \centering
     \includegraphics[width=1\linewidth]{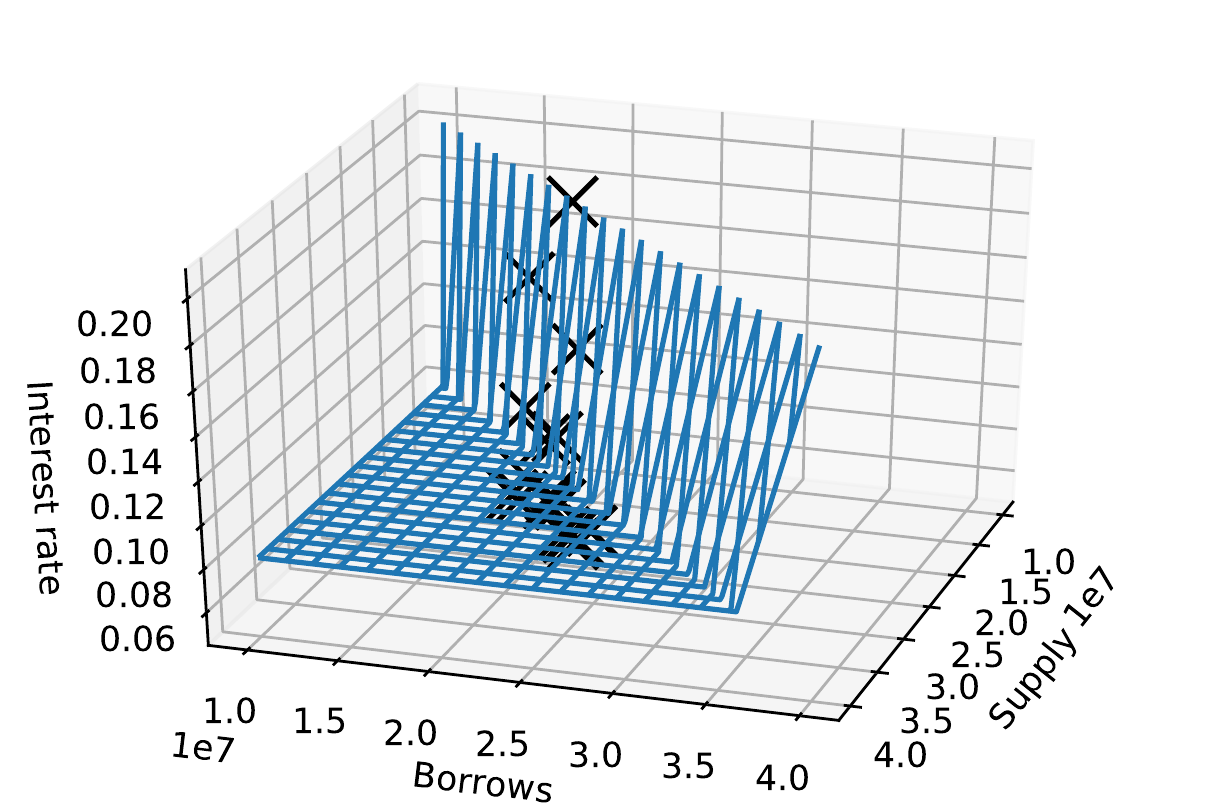}
     \caption{21 February - 13 March}
     \label{fig:compound-wireframe-6}
  \end{subfigure}
  
  \begin{subfigure}[b]{0.47\textwidth}
     \centering
     \includegraphics[width=1\linewidth]{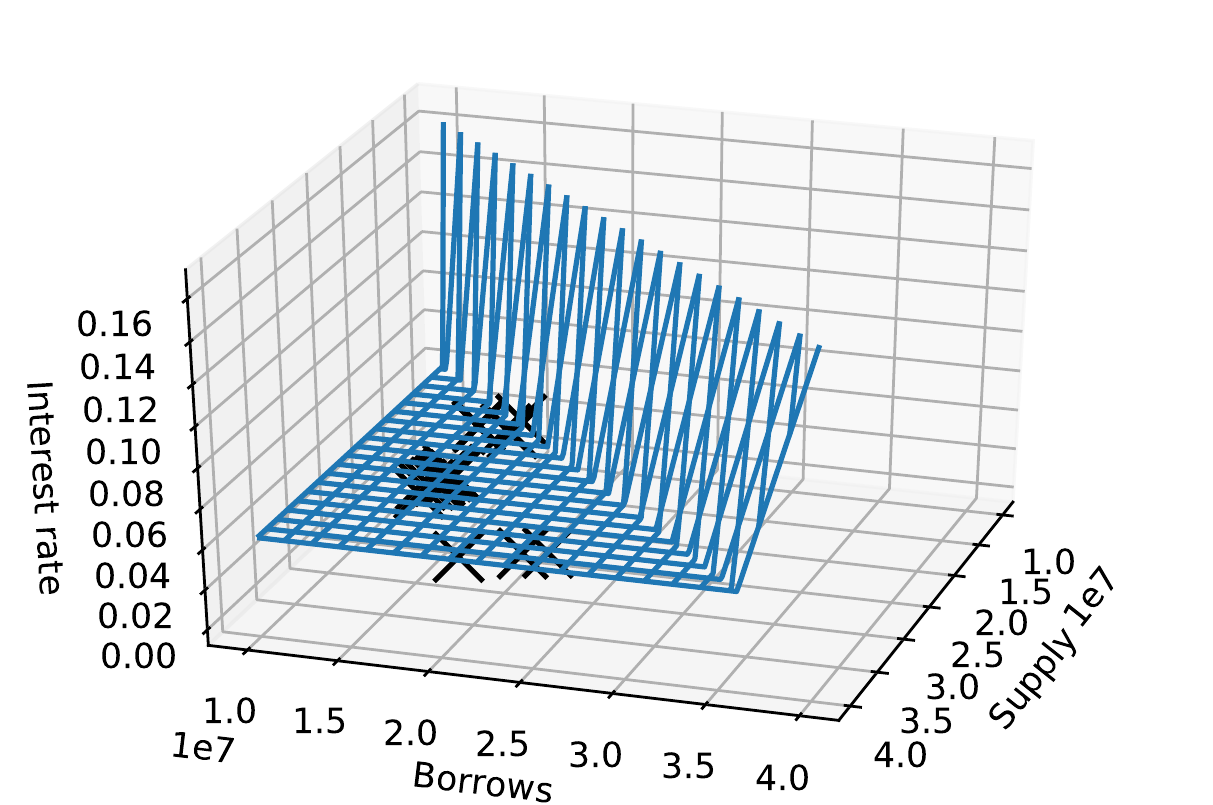}
     \caption{14 March - 5 April}
     \label{fig:compound-wireframe-7}
  \end{subfigure}
  
  \begin{subfigure}[b]{0.47\textwidth}
    \centering
    \includegraphics[width=1\linewidth]{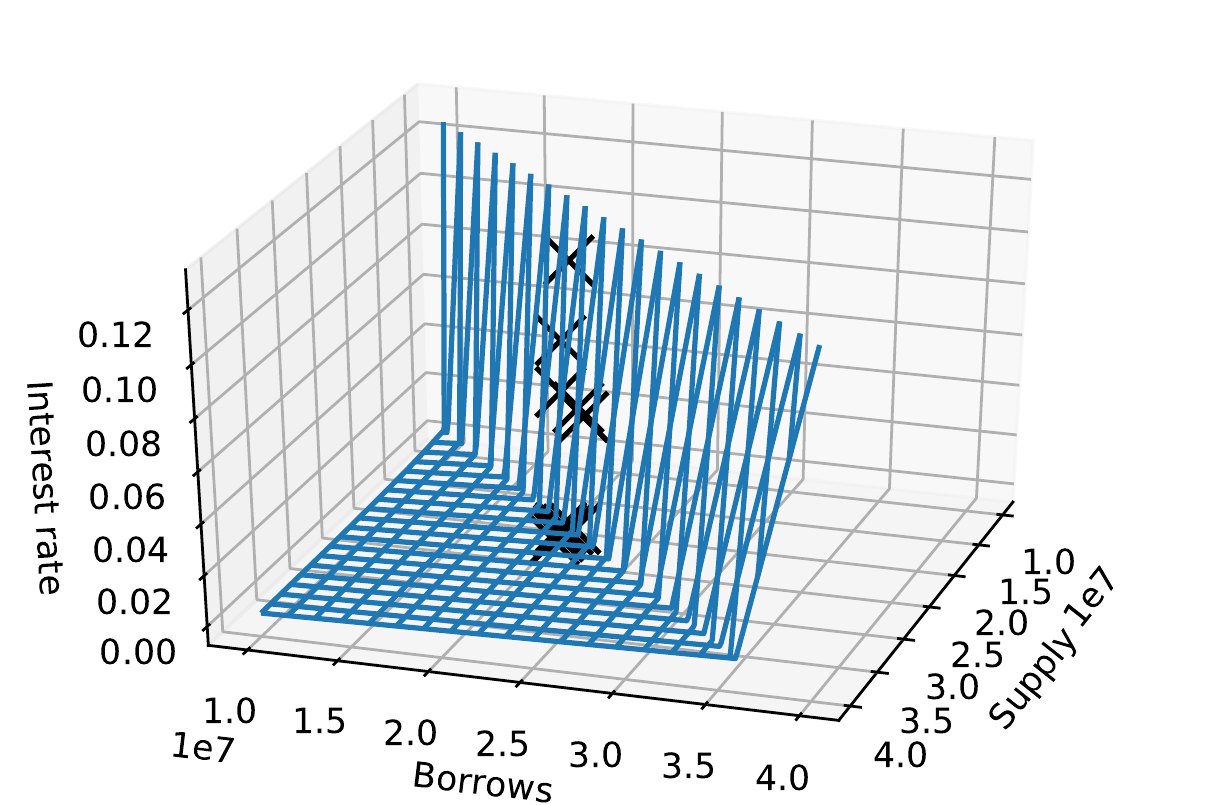}
    \caption{6 April - 21 April}
    \label{fig:compound-wireframe-8}
  \end{subfigure}
  
  \caption[]{Borrowing rates surface for DAI.}
  \label{fig:borrows-surface}
\end{figure}

\begin{figure}[tbp]
  \centering
  \includegraphics[width=\columnwidth]{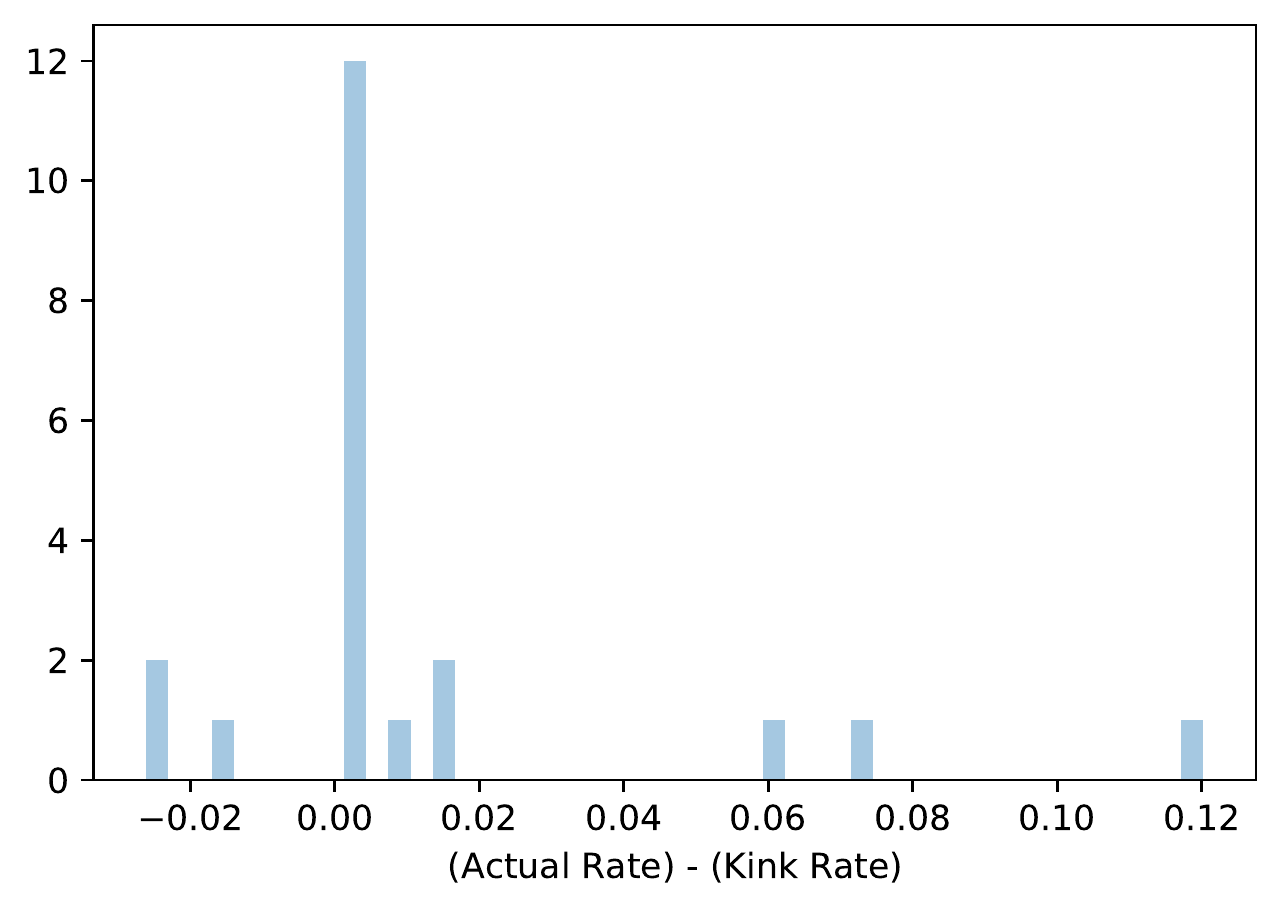}
  \caption{Borrowing rate distribution around kink, 21 February - 13 March.}
  \label{fig:compound-hist-6}
\end{figure}

Figure~\ref{fig:compound-wireframe-6} and the corresponding Figure~\ref{fig:compound-hist-6} plot the interest rate model (the blue surface) as well as the realized interest rate (red crosses).
The two points to note are that (i) there appears to be a clustering of the realized interest rates at the kink of the interest rate function and (ii) otherwise, interest rates are typically higher than the kink, corresponding to a utilization of above 90\%.

\begin{figure}
  \centering
  \begin{subfigure}[b]{0.47\textwidth}
     \centering
     \includegraphics[width=1\linewidth]{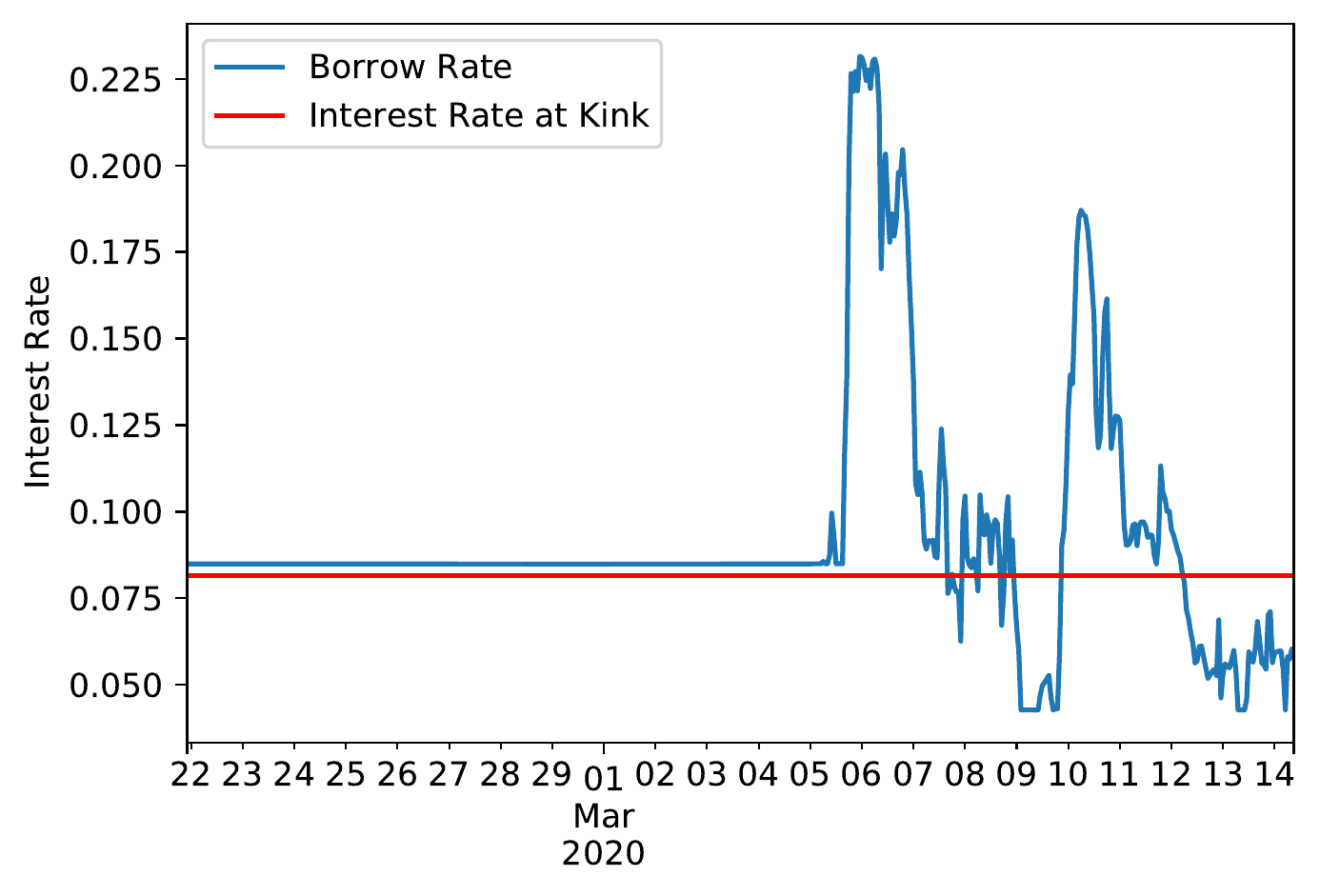}
     \caption{21 February - 13 March}
     \label{fig:compound-rate-vs-kink6}
  \end{subfigure}
  
  \begin{subfigure}[b]{0.47\textwidth}
     \centering
     \includegraphics[width=1\linewidth]{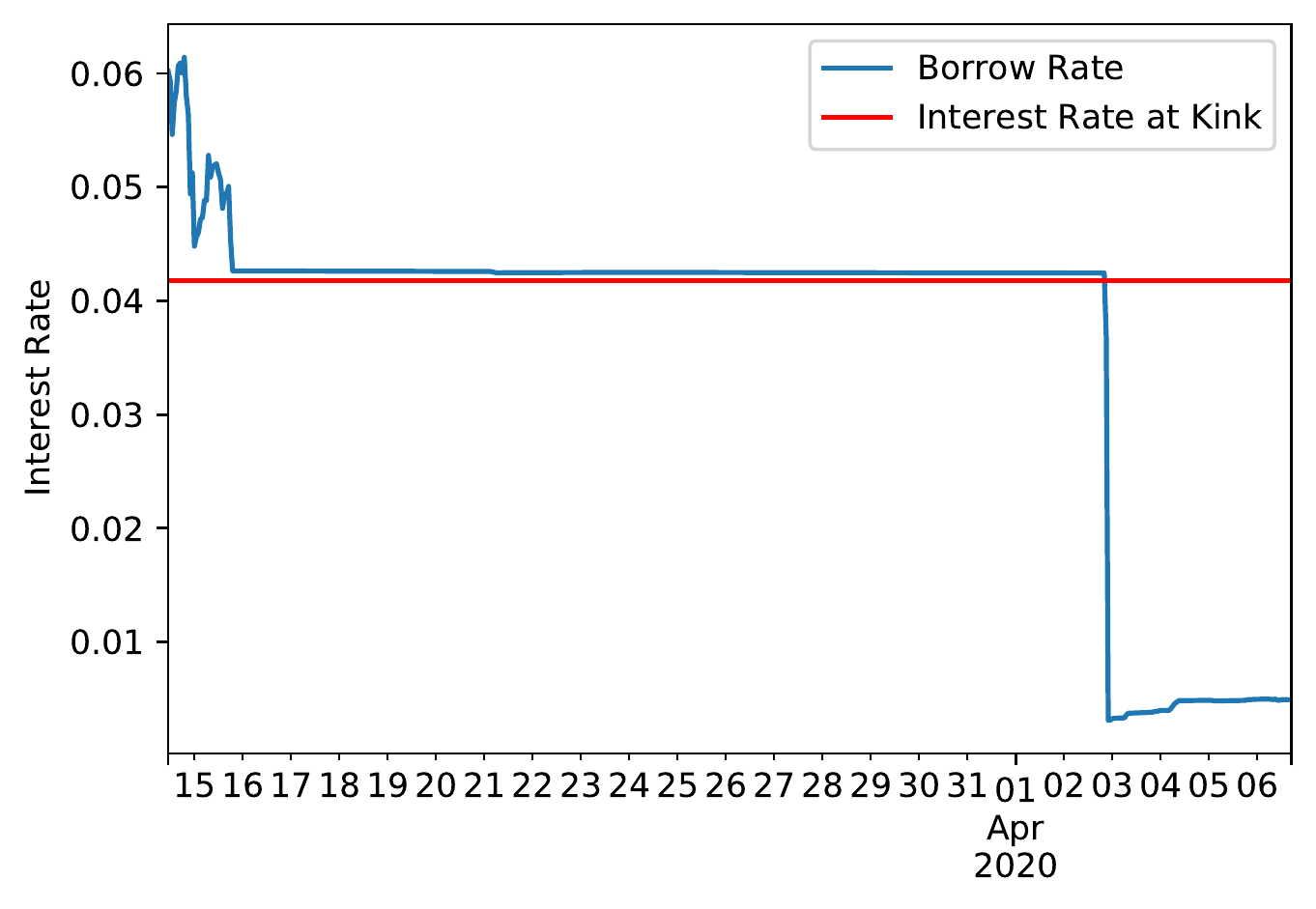}
     \caption{14 March - 5 April}
     \label{fig:compound-rate-vs-kink7}
  \end{subfigure}
  
  \begin{subfigure}[b]{0.47\textwidth}
    \centering
    \includegraphics[width=1\linewidth]{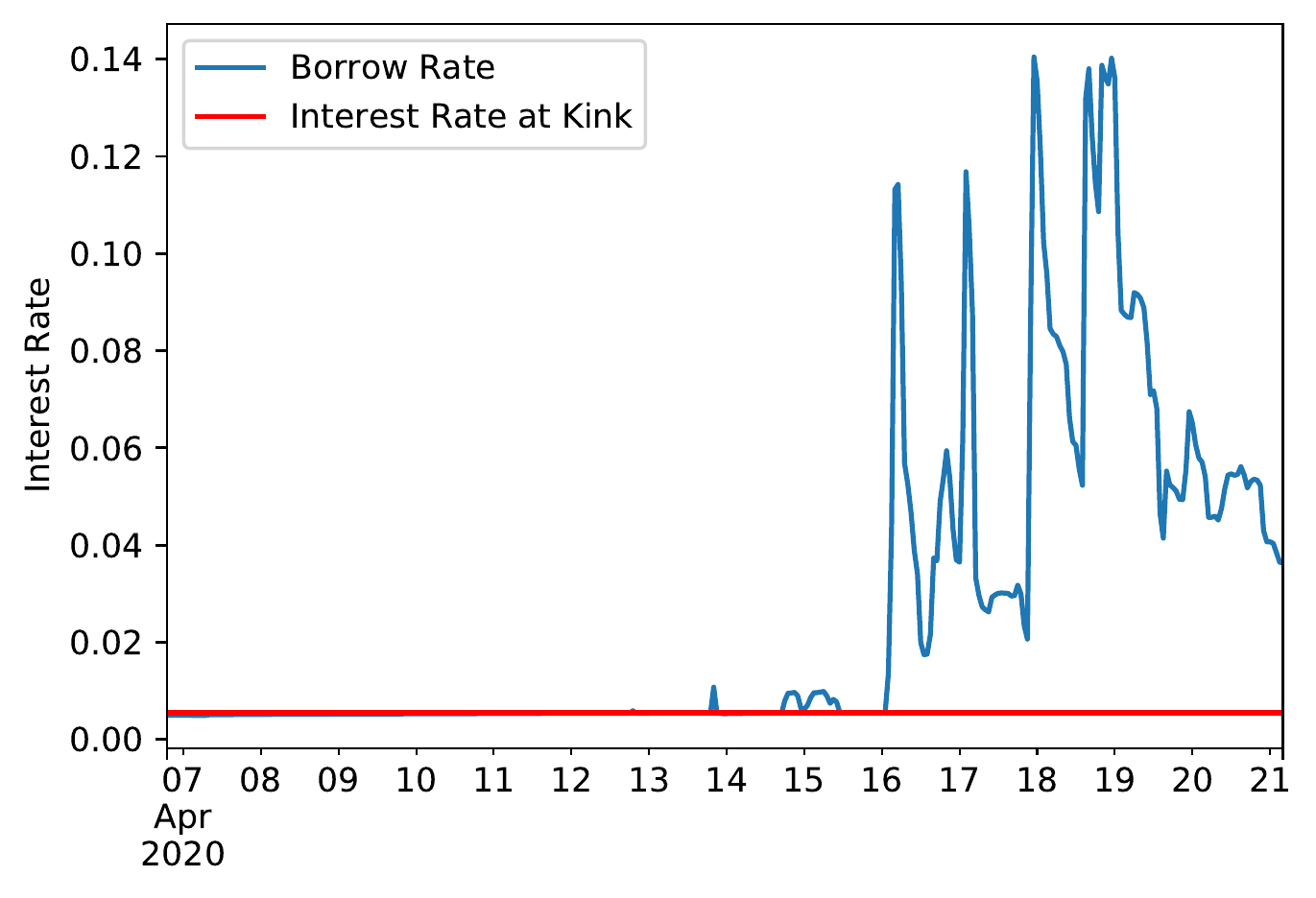}
    \caption{6 April - 21 April}
    \label{fig:compound-rate-vs-kink8}
  \end{subfigure}
  
  \caption[]{Borrowing rate (hourly mean) distribution around kink for DAI. }
  \label{fig:compound-rate-vs-kink}
\end{figure}

Figure~\ref{fig:compound-wireframe-7} shows the interest rate model and the realized interest rates in the next period, after the base rate $\alpha$ is reduced via a parameter change by 49.04\%.
Despite this change, we continue to observe a clustering of the realized interest rates at the kink, although there does appear to be some effect of reducing the typical utilization ratio to below the kink.



Figure~\ref{fig:compound-wireframe-8} shows how the system behaves once the base rate $\alpha$ is set to zero, while the multiplier $\beta$ is increased by nearly 1000\%. 
Again, we observe a similar pattern: most of the realized interest rates appear to be at the kink.
However, if not at the kink, now typically utilization is above 90\%.




\subsection{Summary}
We saw that, especially for DAI, there were several periods of illiquidity and that they were often shared across the three protocols.
We also showed that the locked funds were very concentrated and that a very small amount of accounts had the potential to make the markets illiquid.
Finally, we analyzed the interest rate behavior of DAI on Compound and showed that during all the observation periods, the interest rates appeared clustered around the kink of the interest rate function.

\section{Market Efficiency}
\label{sec:market-efficiency}

In this section we consider the capital market efficiency of DeFi lending protocols.
Loosely, a capital market is said to be efficient if in the process of determining prices, it fully and correctly reflects all relevant information~\cite{malkiel1989efficient}.
More precisely:

\begin{definition}[Market efficiency]
    A market is efficient with respect to some information set $\phi$ if prices would be unaffected by revealing that information to all market participants ~\cite{malkiel1989efficient}. 
\label{def:market-efficiency}    
\end{definition}

A notable consequence of Definition \ref{def:market-efficiency} is that such efficiency implies it the impossibility of making economic profits on the basis of the information set $\phi$.
The market efficiency of PLFs is a question of central interest because it provides a mechanism to assess the maturity of the markets and to understand the responsiveness of agents to changes in the information set $\phi$. 
Moreover, since a core mechanism common to many PLFs is the use of high interest rates at times of high utilization---to encourage saving and discourage borrowing, incentivizing agents to behave in a certain way---the extent to which PLFS are capital efficient will inform how reliable this mechanism is, at present, in incentivizing agents to act in the intended way.
If agents do not in fact respond to high interest rates by reducing their borrowing requirements and increasing their supply of funds to a PLF, illiquidity resulting from high utilization rates on a given protocol may be expected to result. 
Such illiquidity events, where agents cannot withdraw their funds, can be expected to cause panic in financial markets. 
Therefore from the point of view of financial risk, the efficiency of markets is of central interest.

Thus in this section we consider whether PLFs are efficient within a given protocol, considering Compound~\cite{web:compoundfinance} within a framework which is standard in assessing the efficiency of markets in the context of foreign exchange: Uncovered Interest Parity. 

\subsection{Uncovered Interest Parity}
First, we set out Uncovered Interest Parity (UIP) as it would normally appear in the context of foreign exchange between two countries: \emph{domestic} and \emph{foreign}.
An investor has the choice of whether to hold domestic or foreign assets. 
UIP is a theoretical no-arbitrage condition, which states that in equilibrium, if the condition holds, a risk-neutral investor should be indifferent between holding the domestic or foreign assets because the exchange rate is expected to adjust such that returns are equivalent. 

For example, consider UIP holding between GBP and USD. 
An investor starting with 1m GBP at $t=0$ could either:

\begin{itemize}
    \item receive an annual interest rate of $i_{\text{GBP}} = 3\%$, resulting in 1.03m GBP at $t=1$
    \item or, immediately buy 1.23m USD at an exchange rate $S_{\text{GBP/USD}} = 0.8130$,  receiving an annual interest rate of $i_{\text{USD}} = 5\%$, resulting in 1.2915m USD at $t=1$. Once converted to GBP at the new exchange rate at $t=1$, $S_{\text{GBP/USD}} = 0.7974$, identically yields 1.03m GBP.
\end{itemize}

If UIP holds, despite the higher interest rate adjustments in the exchange rate between the currencies offset any potential gain such that arbitrage is not possible. 
Mathematically, UIP is stated as follows. 

\begin{equation}
1 + \iota_i = (1+\iota_j)\frac{\mathrm{E}_t[S_{t+k}]}{S_t}
\label{eqn:uip}
\end{equation}

where $\mathrm{E}_t[S_{t+k}]$ denotes the expectation in period $t$ of the exchange rate $S_{i/j}$ between assets $i$ and $j$ at time $t+k$, $k$ is an arbitrary number of periods into the future, $S_t$ is the current spot exchange rate between assets $i$ and $j$, $\iota_i$ is the interest rate payable on asset $i$ and $\iota_j$ is the interest rate payable on asset $j$.
If Equation (\ref{eqn:uip}) holds, then investors cannot make risk free profit. 

\subsection{UIP in a PLF}
Here, analogously, we perform a pairwise analysis of all possible pairs of tokens available within a protocol, seeking to establish whether UIP holds for that pair.
For UIP to hold it must be the case that a risk-neutral investor would be indifferent between saving (or borrowing) either of the tokens within the pair, because the exchange rate between any token pair adjusts such that no risk-free profit can be made.
As it is the largest PLF~\cite{defipulse} at the time of writing, we consider to what extent the condition holds within Compound~\cite{web:compoundfinance}.

\subsection{Empirical approach}

To develop our empirical specification, we assume that agents have rational expectations: 

\begin{equation}
    S_{t+k} = \mathrm{E_t}[S_{t+k}] + \epsilon_{t+k}
\label{eqn:ratex}
\end{equation}

\noindent where $\epsilon$ denotes a random error. 
Taking logs of equation \ref{eqn:uip} and approximating $\log(1+\iota_i) \approx \iota_i$, we test whether UIP obtains with the following empirical specification:

\begin{equation}
    s_{t+1}-s_t = \alpha + \beta (\iota_i - \iota_j) + \epsilon
\label{eqn:uip-specifiation}
\end{equation}

where $\log{S_{t+1}} = s_{t+1}$ and $\log{S_{t}} = s_{t}$.

\begin{equation}
    \text{\HNull~\textbf{Strict form UIP:}}\quad \alpha=0~\text{and}~ \beta = 1
\label{hyp:a} 
\end{equation}

\noindent Alternatively, we could impose no restriction on $\alpha$ perhaps reflecting a risk premium~\cite{alexius2001uncovered}.

\begin{equation}
    \text{\HNullB~\textbf{Weak form UIP:}}\quad \beta = 1
\label{hyp:b} 
\end{equation}

The existence of a risk premium reflects the extra return in the form of interest payment is required in order for investors to receive the same risk-adjusted return as on a less risky token. 
We test both hypotheses, considering all possible token pairings on Compound and reporting borrowing and saving interest rates separately.

\subsection{UIP regression results - borrowing rates}

For both borrowing and saving rate regressions, we use heteroskedasticity and autocorrelation robust standard errors, which we report in brackets in the results\footnote{WBTC was excluded from the analysis due to data quality issues.}.
Considering first data at the daily frequency (Appendix Table~\ref{tab:uip-daily-borrows}), we find that for 20 market pairs, we reject both~\HNull and~\HNullB at the 1\% level, suggesting UIP does not hold in either its strong or weak form for daily data. 
At the weekly frequency, in Appendix Table~\ref{tab:uip-weekly-borrows}, the evidence is more mixed.
We continue to reject ~\HNull at the 1\% level for 12/20 pairs, but find evidence consistent with ~\HNull for 8 pairs.
Regarding ~\HNullB, we are unable to reject it at the 1\% level in 11 cases.
However, the standard errors are typically large, such that it is difficult to reject any null hypothesis.
Overall, for daily data we find no evidence supportive of UIP holding, while for weekly data we find some evidence that is supportive.

\subsection{UIP regression results - saving rates}

Looking at saving rates at the faily frequency, Appendix Table~\ref{tab:uip-daily-saves}, similarly to borrowing rates for all 20 pairs we reject both~\HNull and~\HNullB at the 1\% level, suggesting UIP does not hold in either its strong or weak form for daily data.
At the weekly frequency, in Appendix Table~\ref{tab:uip-weekly-saves}, the evidence is more mixed.
We continue to reject ~\HNull at the 1\% level for 12/20 pairs, but find evidence consistent with ~\HNull for 8 pairs.
Regarding ~\HNullB, we are unable to reject it at the 1\% level in 9 cases.
However, again the standard errors are typically large, such that it would be difficult to reject any hypothesis.
Overall, for daily data we again find no evidence supportive of UIP holding, while for weekly data we find some evidence that is supportive.

\subsection{Summary}
Looking at daily and weekly frequency data for borrowing and saving, we find weak evidence that UIP holds as the time horizon increases.
This parallels empirical results in traditional foreign exchange markets~\cite{isard2006uncovered}.
This therefore suggests that overall the markets within the Compound PLF may not be fully capital efficient at present, and it seems plausible that these results are not only idiosyncratically true of Compound.
The finding that this PLF is not capital efficient at the daily frequency is not surprising - there is considerable of evidence that UIP does not hold even in traditional foreign exchange markets~\cite{chinn2005testing}.
In addition, this suggests that the \emph{currency carry trade}---where an investor borrows a low yield currency to obtain a high yield currency---is likely to be profitable, since in such inefficient markets differences in yield are not offset by corresponding changes in the exchange rate between the currencies. 
Moreover, we submit that in the context of a PLF, to the extent that there is market inefficiency, agents may not be fully responding to these incentives.

\section{Market dependence}
\label{sec:market-dependence}

We now consider the extent of inter-connectedness \textit{between} protocols by considering how changes in an interest rate for a given token on one PLF are related to changes in the interest rate for the token on another PLF.

For example, consider the borrowing rate for DAI, $i_{b,Dai}$. 
A priori, we would expect that if $i_{b,Dai}$ is higher on one PLF than others, agents would be incentivized to borrow from those PLFs with a lower borrowing rate, deleveraging on one PLF and leveraging on others.
But this influx of borrowers for the token on other PLFs would, in turn, increase the borrowing rates on those protocols. 

In this section, taking the stablecoins DAI and USDC, we investigate whether there is evidence of such dynamics, and find that such behavior is indeed observable. 
Moreover, we quantify the speed of adjustment to new equilibria values, and in so doing measure in one way the responsiveness of agents to their incentives in PLFs. 

\subsection{Vector Error Correction Models}

We model both the short and long run dynamics between borrowing rates for DAI and USDC by using a Vector Error Correction Model (VECM).
Details of this approach are outlined Appendix~\ref{sec:appendix-vecm}.
\subsection{Results}
\begin{figure}
    \centering
    \includegraphics[width=0.42\textwidth]{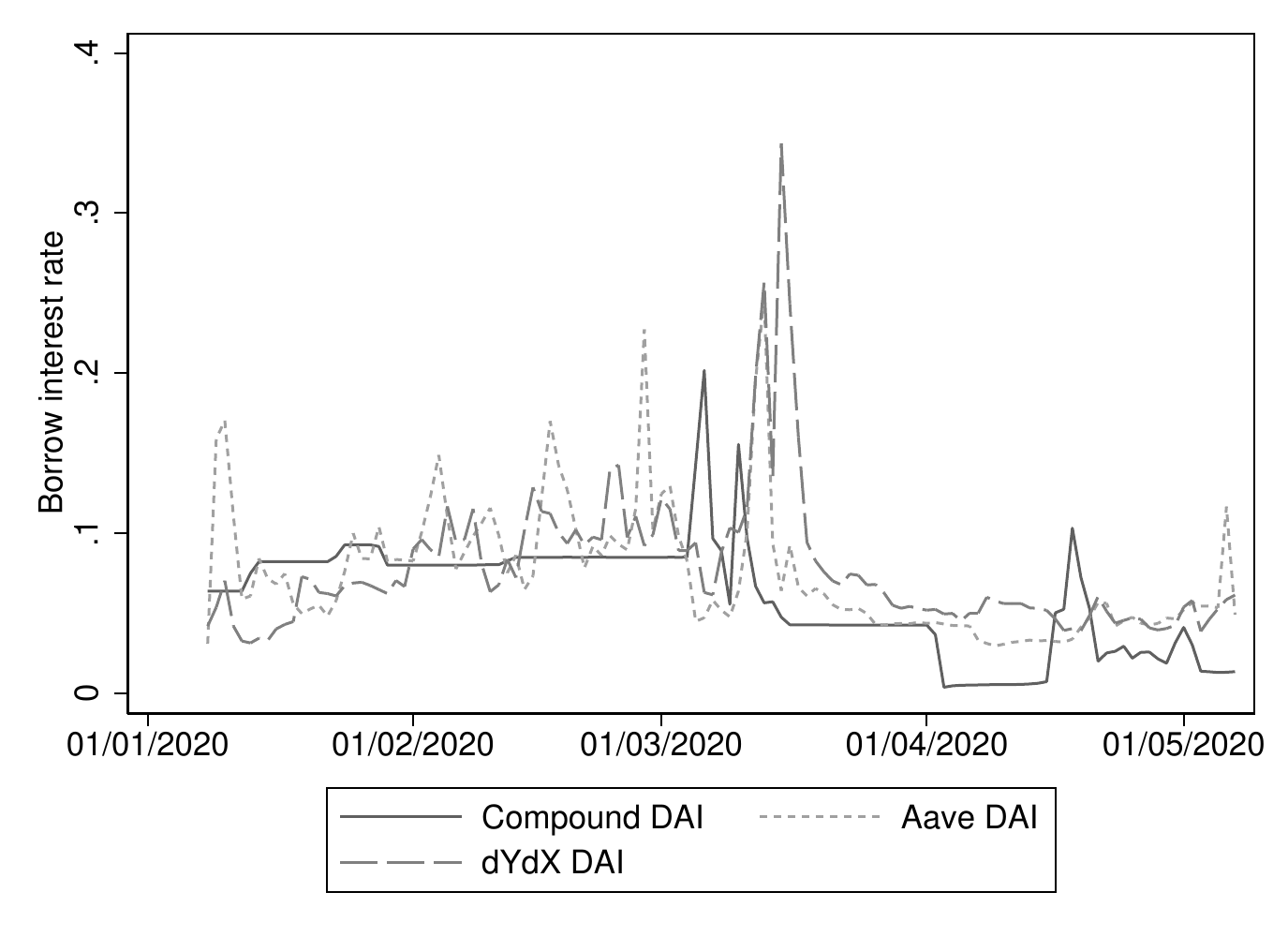}
    \caption{Daily borrowing interest rates on DAI across protocols.}
    \label{fig:between-dai}
\end{figure}

\begin{figure}
    \centering
    \includegraphics[width=0.42\textwidth]{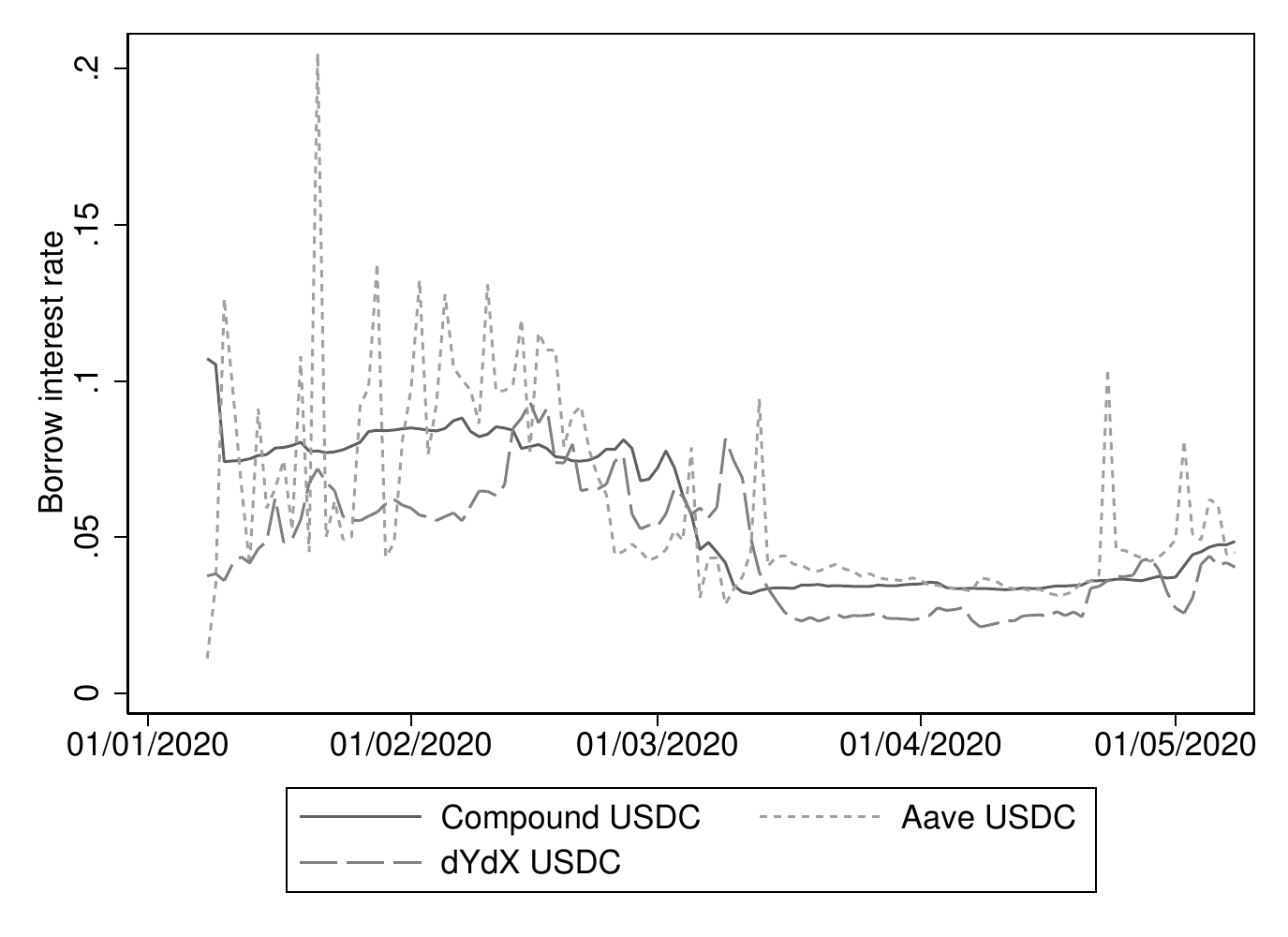}
    \caption{Daily borrowing interest rates on USDC across protocols.}
    \label{fig:between-usdc}
\end{figure}

Separately, we focus on the borrowing rates for DAI and USDC separately, considering Compound, Aave and dYdX. 
We present the borrowing rates for DAI in Figure~\ref{fig:between-dai} and for USDC in Figure~\ref{fig:between-usdc}.

\paragraph{DAI Results}
First, we consider the markets for DAI.
Testing for the number of cointegrating relationships between the three series using Johansen's multiple trace test method~\cite{statacorp2013stata}, we find evidence of at most two cointegrating relationships.
After iteratively tuning the model with postestimation results, we find the optimum lag length to be 5. 
The results are presented in full in Appendix Table~\ref{tab:dai-vecm}.

In terms of short adjustment coefficients, we find a statistically significant coefficient on Aave DAI of $0.38$, such that when the borrowing rate on Compound is high, Aave's borrow rate quickly increases to match it.
Similarly, we find a similar effect for dYdX DAI, this time with an slightly slower adjustment speed of $0.28$.
Interestingly, we do not find evidence of the Compound DAI rate adjusting to changes in the Aave or dYdX DAI rates, suggesting that Compound's interest rate changes drive changes in both Aave and dYdX's borrowing rates, which may suggest that Compound has market power.
This is perhaps to be expected: as we show in Fig.~\ref{fig:dai-liquidity}, Compound has the largest borrow and supply volumes for DAI compared to the other two PLFs and thus will plausibly shape interest rates across protocols.
We obtain the following long-run cointegrating relationships:

\begin{equation}
    \label{eqn:dai-lr-1}
    {DAI}_{Compound} = -1.151 {DAI}_{dYdX} - 0.030
\end{equation}

and

\begin{equation}
    \label{eqn:dai-lr-2}
    {DAI}_{Aave} = -0.991 {DAI}_{dYdX} -0.005
\end{equation}

such that for DAI, dYdX has a long run cointegrating relationship with Compound and Aave. 

\begin{figure}
    \centering
    \includegraphics[width=0.42\textwidth]{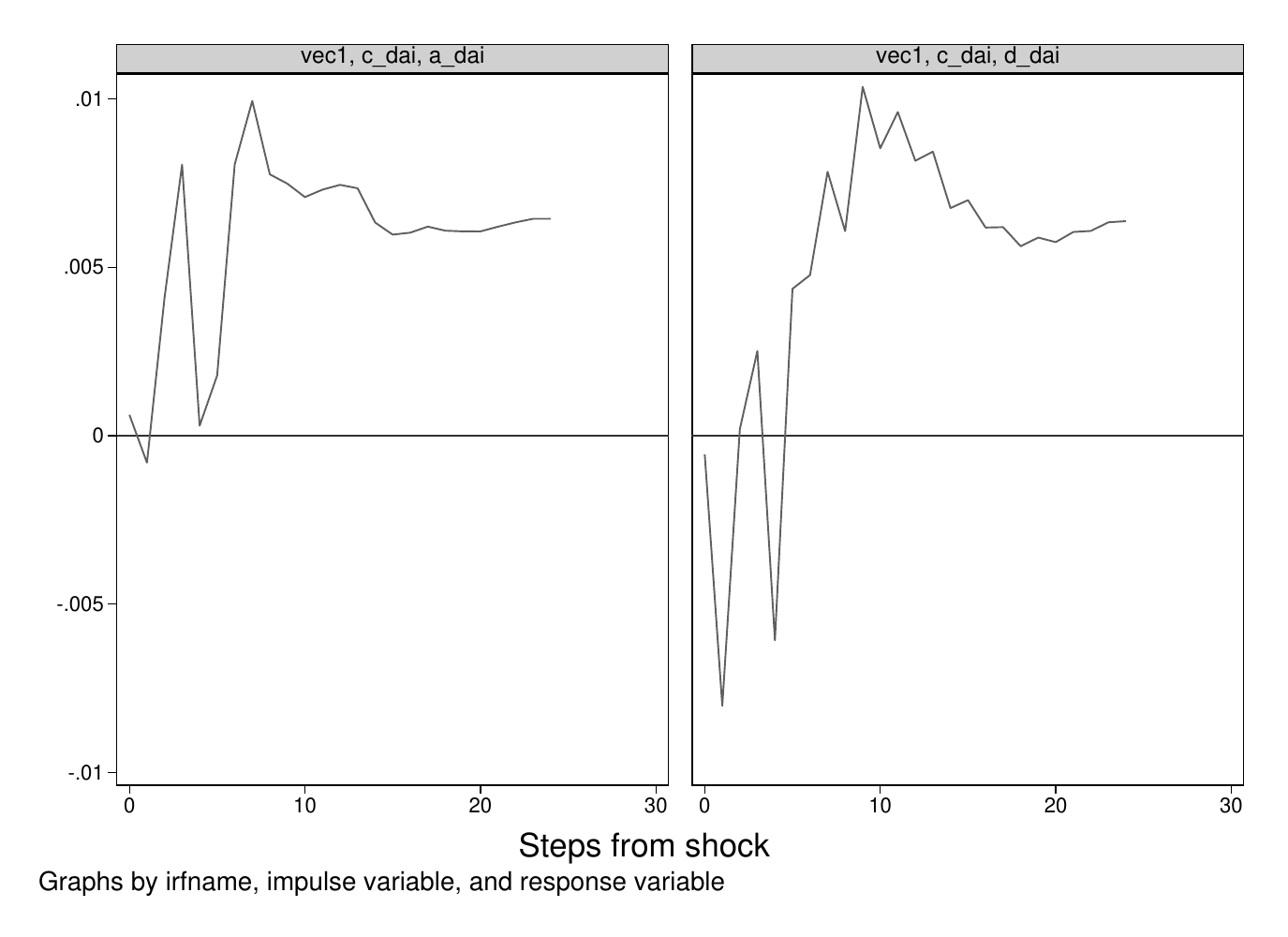}
    \caption{Impulse Response Function: impact of a shock to Compound's DAI borrow rate on Aave and dYdX's DAI borrow rate.}
    \label{fig:irf-dai}
\end{figure}

We present the impact of a shock to Compound's DAI borrow rate on Aave and dYdX's in Figure~\ref{fig:irf-dai}.
It can be seen that a positive shock to the borrowing rate results in a permanent increase in the borrowing rate on Aave and dYdX. 

\paragraph{USDC Results}
For USDC, we find that between the series there are 2 cointegrating relationships~\cite{statacorp2013stata}.
Again, testing for the number of cointegrating relationships between the three series using Johansen's multiple trace test method~\cite{statacorp2013stata}, we find evidence of at most two cointegrating relationships.
After iteratively tuning the model with postestimation results, we find the optimum lag length to be 3.
The results are presented in full in Appendix Table~\ref{tab:usdc-vecm}.

It appears that again Compound has market power, with the borrowing rates on Aave and dYdX adjusting to match the Compound interest rate level.
Aave appears to adjust with a faster speed of $0.607$, in comparison to dYdX at $0.115$.
In terms of long-run relationships, we find that Compound and dYdX share a long-run relationship, and that Aave and dYdX share a long-run relationship.
We obtain the following long-run cointegrating relationships:

\begin{equation}
    \label{eqn:usdc-lr-1}
    {USDC}_{Compound} = -1.353 {USDC}_{dYdX} - 0.007
\end{equation}

and

\begin{equation}
    \label{eqn:usdc-lr-2}
    {USDC}_{Aave} = -1.347 {USDC}_{dYdX} -0.003
\end{equation}

such that for USDC, similarly to DAI, dYdX has a long run cointegrating relationship with Compound and Aave. 

\begin{figure}
    \centering
    \includegraphics[width=0.43\textwidth]{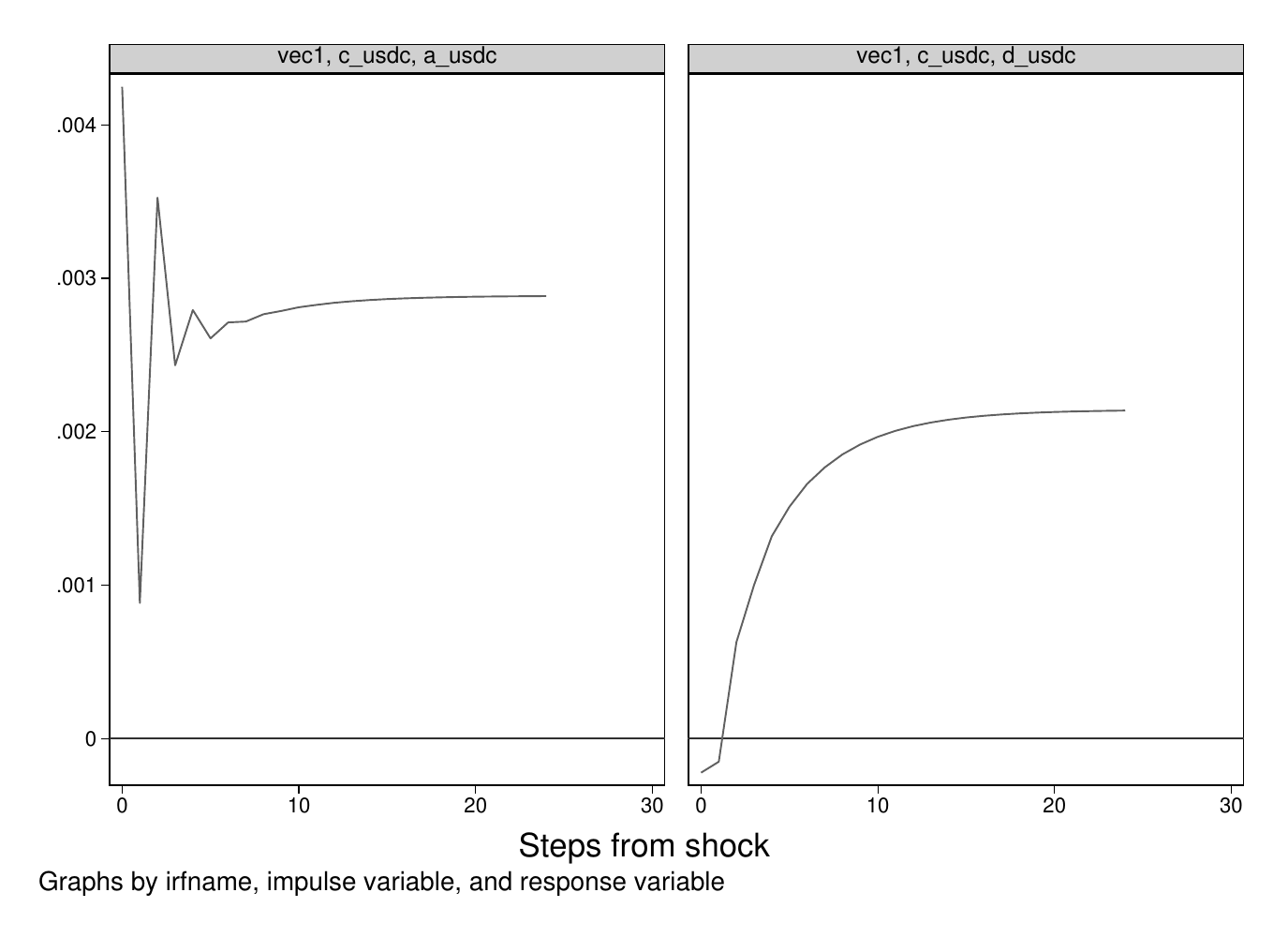}
    \caption{Impulse Response Function: impact of a shock to Compound's USDC borrow rate on Aave and dYdX's USDC borrow rates.}
    \label{fig:irf-usdc}
\end{figure}

We plot the impact of a change in the USDC borrowing rate in Figure~\ref{fig:irf-usdc}.
A shock to Compound's borrowing rate on USDC has a permanent effect on the interest rates in Aave and dYdX. 

\subsection{Robustness checks}
We performed extensive robustness checks on the fitted VECM models. 
Since our ability to draw sound inference on the adjustment parameters depends on the cointegrating equations being stationary, we plot the cointegrating equations over time (see Appendix Figures~\ref{fig:coint-dai1},~\ref{fig:coint-dai2}, ~\ref{fig:coint-usdc1} and \ref{fig:coint-usdc2}.)
We argue that the cointegrating equations appear overall without significant trends, though note the presence of a large negative shock in the DAI specifications mid-March, and therefore broadly stationary.
Furthermore, we check that we have correctly specified the number of cointegrating equations in Appendix  Figures~\ref{fig:dai-vecstable} and ~\ref{fig:usdc-vecstable}.
We find no evidence that any of the eigenvalues are close to the unit circle, and therefore no evidence that the model is misspecified (see ~\cite{statacorp2013stata} for details on this test.)
Additionally, we test for serial correlation in the residuals of the regressions and find little evidence of this.
A test for the normality of the errors in our models does suggest that the errors are non-normally distributed, which may affect our standard errors but should not result in parameter bias. 
Jointly this panel of robustness tests gives us confidence that the VECM models are reasonably well specified.

\subsection{Summary}
Overall we find evidence of cointegrating relationships between markets for DAI and USDC. 
In turn, this suggests that to some extent interest rate changes in one protocol are associated with interest rate changes in others, perhaps in turn providing evidence of agents being incentivized to change protocol by the rates they observe.
Moreover, we also find some evidence of Compound having market power.

\section{Related Work}
\label{sec:related_work}

In this section, we present related work about interest rates in both traditional finance as well as in DeFi protocols.
Since, to the best of our knowledge, this paper is the first academic work to investigate PLFs in detail, we include some non-academic work which covers some aspects of PLFs interest rates.

The authors of~\cite{kao2020analysis} focus on the Compound protocol and present an overview of the market risks and liquidation mechanism. 
They perform agent-based simulations to investigate the economic security of the protocol, and find that the protocol is able to scale to a larger market size while maintaining a low probability even when markets are volatile. 

In~\cite{lending-pool-interests}, the author describes how the interest rate models work in PLFs.
The author first provides a definition of the utilization ratio of a PLF, then describes linear and polynomial interest rate models and finally presents how these different models are used by three major PLFs, namely, Compound, dYdX and DDEX~\cite{ddex}.

The author of~\cite{illiquidity-defi} analyzes Compound to show the risks inherent to decentralized lending.
In particular, they focus on the risks associated with illiquidity and bank runs.
The authors analyze the SAI market on compound and find that there were several periods of near-illiquidity and actual-illiquidity.
They present instances where the illiquidity is created because of large loans in a short period of time and others where it is created by the lenders withdrawing large amount of funds they had locked.
In particular, they show that on five occasions, a single transaction was sufficient to withdraw more than a quarter of the available liquidity, and in the worst case a single transaction drained more than 95\% of the available liquidity.

The author of~\cite{aave-black-thursday} focuses on how Black Thursday~\cite{black-thursday} in March 2020 affected the Aave market.
They first show that the amount of money borrowed through flash loans went up by more than 10,000\% in only a few hours because users were leveraging these to liquidate their collateralized debt positions~\cite{web:maker,whitepaper:maker}.
The author also highlights the fact that the amount of borrows liquidated on Aave during Black Thursday was more than 100 times higher than the typical amount liquidated, reaching a total of more than 550k USD in a single day.
Finally, the author show that during the Black Thursday crisis, some design flaws of MakerDAO's protocol~\cite{whitepaper:maker} caused Maker to loose a total of more than 4m USD worth of collateral.

In~\cite{Jakab2015}, the authors elucidate the difference between the intermediation and financing roles of traditional banks, and show that when modelling banks with financing models as opposed to intermediation models, identical shocks have much greater effects on the real economy. 

Finally,~\citet{brody2019theory} present a work about interest rates in the context of cryptocurrencies but centered on a different problem.
The focus of their work lies on how cryptocurrencies could set an interest rate for their holders, such that that they accumulate these interest rates in a continuous manner.





\section{Conclusion}
\label{sec:conclusion}

In this paper, we coin the phrase Protocol for Loanable Funds, to describe DeFi equivalents of Intermediaries for Loanable Funds in traditional finance, providing a classification framework for the extant interest rate models. 
We analyze three of the largest PLFs in terms of market liquidity, efficiency and dependence.

In terms of market liquidity we find we find that individually PLFs often operate at times of high utilization, and moreover, often these moments of high utilization are shared across protocols.
Moreover, we find that token holdings can be concentrated to a very small set of accounts, such that at any time were a small number of suppliers to withdraw their funds, perhaps in concert, they could significantly reduce the liquidity available on markets and perhaps make such markets illiquid.

In terms of market efficiency, we consider whether uncovered interest parity holds.
On the whole, we find that it does not, suggesting that token markets are at present relatively inefficient. 
This also suggests that at present agents may not be fully responsive to interest rate incentives.

In terms of market dependence we find  that the borrowing rates on these protocols influence each other, an in particular that Compound appears to have some market power to set the prevailing borrowing rate for Aave and dYdX.

\begin{acks}
We thank the anonymous reviewers for their feedback and suggestions.
This project received partial funding from EPSRC Standard Research Studentship (DTP) (EP/R513052/1), the Tezos Foundation and the Brevan Howard Centre for Financial Analysis. 
\end{acks}

\balance
\bibliographystyle{ACM-Reference-Format}

\bibliography{references.bib}
\appendix
\clearpage
\onecolumn
\section{Appendix}
\label{sec:appendix}

\setcounter{table}{0}
\setcounter{figure}{0}
\renewcommand{\thetable}{A\arabic{table}}
\renewcommand{\thefigure}{A\arabic{figure}}

\begin{table*}[h]
    \centering
    \begin{tabular}{l r r r r}
    \toprule
    & \multicolumn{4}{c}{Parameters} \\
    \cline{2-5}
    Date & $\alpha$ & $\beta$ & $\gamma$ & $U^{*}$ \\
    \midrule
    17 Dec '19 & 19637062989 & 264248265 & 570776255707 & 9e17 \\
    8 Jan '20 & 29174130900 & 264248265 & 570776255707 & 9e17 \\
    26 Jan '20 & 37372598273 & 264248265 & 570776255707 & 9e17 \\
    4 Feb '20 & 41997859121 & 264248265 & 570776255707 & 9e17 \\
    9 Feb '20 & 36209575847 & 705029680 & 570776255707 & 9e17 \\
    21 Feb '20 & 38532925389 & 264248265 & 570776255707 & 9e17 \\
    14 Mar '20 & 19637062989 & 264248265 & 570776255707 & 9e17 \\
    6 Apr '20 & 0 & 2900146648 & 570776255707 & 9e17 \\
    21 Apr '20 & 0 & 264248265 & 570776255707 & 9e17 \\
    27 Apr '20 & 0 & 10569930661 & 570776255707 & 9e17 \\
    \bottomrule
    \end{tabular}
    \caption{Interest rate model and parameter changes for the cDAI contract since 17th December 2019 (prior to this date an earlier variation of the interest rate model ---`Jump Rate Model'---was in force since 23rd November 2019; we omit this period for expositional clarity.).}
    \label{tab:cdai-models}
\end{table*}

\clearpage

\begin{table*}
\caption{Table of UIP results for daily frequency data, using borrowing rates.
Using Newey-West heteroscedasticity and autocorrelation robust standard errors (reported in parentheses.)}
\small
\begin{tabular}{lllllllcc}
        \toprule
Pair & N.obs & $\alpha$ & $\beta$ & R-squared & $\alpha$ p-value & $\beta$ p-value  &  Strict form (\ref{hyp:a}) p-value & Weak form (\ref{hyp:b}) p-value \\
\midrule
  eth\_bat &   392 &    0.01 &   -0.482483 &      0.02 &          0.05 &         0.11 &           0.00 &     0.00 \\
          &       &  (0.01) &      (0.30) &           &               &              &                &          \\
  eth\_zrx &   389 &    0.00 &   -0.194958 &      0.00 &          0.30 &         0.17 &           0.00 &     0.00 \\
          &       &  (0.00) &      (0.14) &           &               &              &                &          \\
 eth\_usdc &   393 &    0.00 &  -0.0357526 &      0.00 &          0.74 &         0.76 &           0.00 &     0.00 \\
          &       &  (0.01) &      (0.12) &           &               &              &                &          \\
  eth\_dai &   175 &    0.01 &   -0.252845 &      0.01 &          0.10 &         0.22 &           0.00 &     0.00 \\
          &       &  (0.01) &      (0.20) &           &               &              &                &          \\
  eth\_sai &   397 &    0.00 &  -0.0315418 &      0.00 &          0.61 &         0.58 &           0.00 &     0.00 \\
          &       &  (0.01) &      (0.06) &           &               &              &                &          \\
  eth\_rep &   392 &    0.00 &   0.0512982 &      0.00 &          0.58 &         0.65 &           0.00 &     0.00 \\
          &       &  (0.00) &      (0.11) &           &               &              &                &          \\
  bat\_zrx &   387 &   -0.00 &   -0.478467 &      0.03 &          0.64 &         0.05 &           0.00 &     0.00 \\
          &       &  (0.00) &      (0.25) &           &               &              &                &          \\
 bat\_usdc &   392 &    0.00 &  -0.0913661 &      0.00 &          0.60 &         0.40 &           0.00 &     0.00 \\
          &       &  (0.01) &      (0.11) &           &               &              &                &          \\
  bat\_dai &   175 &    0.00 &   -0.328409 &      0.02 &          0.37 &         0.11 &           0.00 &     0.00 \\
          &       &  (0.00) &      (0.20) &           &               &              &                &          \\
  bat\_sai &   393 &    0.01 &   -0.134668 &      0.01 &          0.20 &         0.11 &           0.00 &     0.00 \\
          &       &  (0.01) &      (0.08) &           &               &              &                &          \\
  bat\_rep &   388 &    0.00 &   0.0854052 &      0.00 &          0.64 &         0.65 &           0.00 &     0.00 \\
          &       &  (0.00) &      (0.19) &           &               &              &                &          \\
 zrx\_usdc &   388 &    0.00 &  -0.0676933 &      0.00 &          0.62 &         0.54 &           0.00 &     0.00 \\
          &       &  (0.01) &      (0.11) &           &               &              &                &          \\
  zrx\_dai &   175 &    0.01 &   -0.514228 &      0.03 &          0.09 &         0.04 &           0.00 &     0.00 \\
          &       &  (0.01) &      (0.25) &           &               &              &                &          \\
  zrx\_sai &   389 &    0.01 &  -0.0759909 &      0.00 &          0.38 &         0.27 &           0.00 &     0.00 \\
          &       &  (0.01) &      (0.07) &           &               &              &                &          \\
  zrx\_rep &   387 &   -0.00 &    -0.23005 &      0.01 &          0.60 &         0.08 &           0.00 &     0.00 \\
          &       &  (0.00) &      (0.13) &           &               &              &                &          \\
 usdc\_dai &   175 &   -0.00 &  -0.0111104 &      0.00 &          0.77 &         0.53 &           0.00 &     0.00 \\
          &       &  (0.00) &      (0.02) &           &               &              &                &          \\
 usdc\_sai &   394 &    0.00 & -0.00474335 &      0.00 &          0.74 &         0.52 &           0.00 &     0.00 \\
          &       &  (0.00) &      (0.01) &           &               &              &                &          \\
 usdc\_rep &   390 &   -0.00 &  -0.0510679 &      0.00 &          0.77 &         0.64 &           0.00 &     0.00 \\
          &       &  (0.01) &      (0.11) &           &               &              &                &          \\
  dai\_sai &   175 &    0.01 &   -0.267694 &      0.01 &          0.24 &         0.08 &           0.00 &     0.00 \\
          &       &  (0.01) &      (0.15) &           &               &              &                &          \\
  dai\_rep &   175 &   -0.01 &   -0.158339 &      0.00 &          0.20 &         0.38 &           0.00 &     0.00 \\
          &       &  (0.00) &      (0.18) &           &               &              &                &          \\
\bottomrule
\end{tabular}
\label{tab:uip-daily-borrows}
\end{table*}

\begin{table*}
\caption{Table of UIP results for daily frequency data, using saving rates. 
Using Newey-West heteroscedasticity and autocorrelation robust standard errors (reported in parentheses.)}
\small
\begin{tabular}{lllllllcc}
        \toprule
Pair & N.obs & $\alpha$ & $\beta$ & R-squared & $\alpha$ p-value & $\beta$ p-value  &  Strict form (\ref{hyp:a}) p-value & Weak form (\ref{hyp:b}) p-value \\
\midrule
  eth\_bat &   392 &    0.00 &    -1.10434 &      0.02 &          0.02 &         0.00 &           0.00 &     0.00 \\
          &       &  (0.00) &      (0.28) &           &               &              &                &          \\
  eth\_zrx &   389 &    0.00 &    -0.61295 &      0.00 &          0.51 &         0.09 &           0.00 &     0.00 \\
          &       &  (0.00) &      (0.36) &           &               &              &                &          \\
 eth\_usdc &   393 &    0.00 &  -0.0164582 &      0.00 &          0.86 &         0.91 &           0.00 &     0.00 \\
          &       &  (0.01) &      (0.15) &           &               &              &                &          \\
  eth\_dai &   175 &    0.01 &    -0.21499 &      0.01 &          0.13 &         0.24 &           0.00 &     0.00 \\
          &       &  (0.01) &      (0.18) &           &               &              &                &          \\
  eth\_sai &   397 &   -0.00 & -0.00335582 &      0.00 &          0.89 &         0.95 &           0.00 &     0.00 \\
          &       &  (0.00) &      (0.05) &           &               &              &                &          \\
  eth\_rep &   392 &    0.00 &  -0.0695602 &      0.00 &          0.44 &         0.28 &           0.00 &     0.00 \\
          &       &  (0.00) &      (0.06) &           &               &              &                &          \\
  bat\_zrx &   387 &   -0.00 &    -1.37843 &      0.05 &          0.76 &         0.00 &           0.00 &     0.00 \\
          &       &  (0.00) &      (0.48) &           &               &              &                &          \\
 bat\_usdc &   392 &    0.00 &   -0.100931 &      0.00 &          0.70 &         0.52 &           0.00 &     0.00 \\
          &       &  (0.01) &      (0.16) &           &               &              &                &          \\
  bat\_dai &   175 &    0.01 &   -0.263267 &      0.01 &          0.11 &         0.13 &           0.00 &     0.00 \\
          &       &  (0.01) &      (0.17) &           &               &              &                &          \\
  bat\_sai &   393 &    0.00 &      -0.095 &      0.01 &          0.35 &         0.14 &           0.00 &     0.00 \\
          &       &  (0.00) &      (0.06) &           &               &              &                &          \\
  bat\_rep &   388 &    0.00 &   0.0467639 &      0.00 &          0.85 &         0.85 &           0.00 &     0.00 \\
          &       &  (0.00) &      (0.25) &           &               &              &                &          \\
 zrx\_usdc &   388 &    0.00 &  -0.0559146 &      0.00 &          0.74 &         0.70 &           0.00 &     0.00 \\
          &       &  (0.01) &      (0.14) &           &               &              &                &          \\
  zrx\_dai &   175 &    0.02 &   -0.341124 &      0.02 &          0.09 &         0.08 &           0.00 &     0.00 \\
          &       &  (0.01) &      (0.19) &           &               &              &                &          \\
  zrx\_sai &   389 &    0.00 &  -0.0586814 &      0.00 &          0.47 &         0.31 &           0.00 &     0.00 \\
          &       &  (0.01) &      (0.06) &           &               &              &                &          \\
  zrx\_rep &   387 &    0.00 &   -0.692352 &      0.01 &          0.96 &         0.01 &           0.00 &     0.00 \\
          &       &  (0.00) &      (0.28) &           &               &              &                &          \\
 usdc\_dai &   175 &    0.00 &  -0.0174017 &      0.00 &          0.31 &         0.44 &           0.00 &     0.00 \\
          &       &  (0.00) &      (0.02) &           &               &              &                &          \\
 usdc\_sai &   394 &    0.00 & -0.00493231 &      0.00 &          0.77 &         0.51 &           0.00 &     0.00 \\
          &       &  (0.00) &      (0.01) &           &               &              &                &          \\
 usdc\_rep &   390 &   -0.00 &  -0.0931906 &      0.00 &          0.67 &         0.50 &           0.00 &     0.00 \\
          &       &  (0.01) &      (0.14) &           &               &              &                &          \\
  dai\_sai &   175 &   -0.00 &   -0.216161 &      0.01 &          0.24 &         0.09 &           0.00 &     0.00 \\
          &       &  (0.00) &      (0.13) &           &               &              &                &          \\
  dai\_rep &   175 &   -0.01 &   -0.153443 &      0.00 &          0.23 &         0.39 &           0.00 &     0.00 \\
          &       &  (0.01) &      (0.18) &           &               &              &                &          \\
\bottomrule
\end{tabular}
\label{tab:uip-daily-saves}
\end{table*}

\begin{table*}
\caption{Table of UIP results for weekly frequency data, using borrowing rates. 
Using Newey-West heteroscedasticity and autocorrelation robust standard errors (reported in parentheses.)}
\small
\begin{tabular}{lllllllcc}
        \toprule
Pair & N.obs & $\alpha$ & $\beta$ & R-squared & $\alpha$ p-value & $\beta$ p-value  &  Strict form (\ref{hyp:a}) p-value & Weak form (\ref{hyp:b}) p-value \\
\midrule
  eth\_bat &    56 &    0.05 &   -2.27281 &      0.02 &          0.47 &         0.56 &           0.64 &     0.40 \\
          &       &  (0.07) &     (3.86) &           &               &              &                &          \\
  eth\_zrx &    56 &    0.01 &  -0.652686 &      0.00 &          0.71 &         0.24 &           0.00 &     0.00 \\
          &       &  (0.02) &     (0.56) &           &               &              &                &          \\
 eth\_usdc &    56 &    0.02 &  -0.254169 &      0.00 &          0.72 &         0.69 &           0.00 &     0.05 \\
          &       &  (0.06) &     (0.64) &           &               &              &                &          \\
  eth\_dai &    25 &    0.07 &   -1.86555 &      0.12 &          0.18 &         0.30 &           0.30 &     0.12 \\
          &       &  (0.05) &     (1.80) &           &               &              &                &          \\
  eth\_sai &    56 &    0.02 &  -0.263629 &      0.01 &          0.52 &         0.47 &           0.00 &     0.00 \\
          &       &  (0.03) &     (0.36) &           &               &              &                &          \\
  eth\_rep &    56 &    0.01 &   0.718818 &      0.01 &          0.68 &         0.48 &           0.91 &     0.78 \\
          &       &  (0.02) &     (1.02) &           &               &              &                &          \\
  bat\_zrx &    56 &   -0.01 &   -1.03856 &      0.01 &          0.44 &         0.46 &           0.10 &     0.15 \\
          &       &  (0.02) &     (1.41) &           &               &              &                &          \\
 bat\_usdc &    56 &    0.03 &  -0.630377 &      0.03 &          0.46 &         0.15 &           0.00 &     0.00 \\
          &       &  (0.04) &     (0.44) &           &               &              &                &          \\
  bat\_dai &    25 &    0.02 &   -2.11941 &      0.17 &          0.49 &         0.12 &           0.10 &     0.03 \\
          &       &  (0.03) &     (1.37) &           &               &              &                &          \\
  bat\_sai &    56 &    0.07 &  -0.920485 &      0.10 &          0.12 &         0.04 &           0.00 &     0.00 \\
          &       &  (0.05) &     (0.45) &           &               &              &                &          \\
  bat\_rep &    56 &    0.03 &    2.65716 &      0.09 &          0.07 &         0.01 &           0.17 &     0.13 \\
          &       &  (0.02) &     (1.09) &           &               &              &                &          \\
 zrx\_usdc &    56 &    0.04 &    -0.5622 &      0.01 &          0.59 &         0.53 &           0.00 &     0.09 \\
          &       &  (0.07) &     (0.90) &           &               &              &                &          \\
  zrx\_dai &    25 &    0.09 &   -3.48611 &      0.17 &          0.18 &         0.08 &           0.06 &     0.03 \\
          &       &  (0.07) &     (1.98) &           &               &              &                &          \\
  zrx\_sai &    56 &    0.05 &  -0.544193 &      0.02 &          0.17 &         0.21 &           0.00 &     0.00 \\
          &       &  (0.04) &     (0.44) &           &               &              &                &          \\
  zrx\_rep &    56 &    0.01 &  -0.702555 &      0.01 &          0.75 &         0.08 &           0.00 &     0.00 \\
          &       &  (0.02) &     (0.40) &           &               &              &                &          \\
 usdc\_dai &    25 &   -0.00 & -0.0976848 &      0.10 &          0.55 &         0.23 &           0.00 &     0.00 \\
          &       &  (0.00) &     (0.08) &           &               &              &                &          \\
 usdc\_sai &    56 &    0.00 & -0.0525398 &      0.09 &          0.08 &         0.02 &           0.00 &     0.00 \\
          &       &  (0.00) &     (0.02) &           &               &              &                &          \\
 usdc\_rep &    56 &   -0.03 &  -0.593887 &      0.03 &          0.29 &         0.08 &           0.00 &     0.00 \\
          &       &  (0.03) &     (0.34) &           &               &              &                &          \\
  dai\_sai &    25 &    0.07 &   -1.84099 &      0.12 &          0.39 &         0.28 &           0.00 &     0.11 \\
          &       &  (0.09) &     (1.69) &           &               &              &                &          \\
  dai\_rep &    25 &   -0.07 &    -1.9174 &      0.16 &          0.10 &         0.14 &           0.10 &     0.03 \\
          &       &  (0.04) &     (1.29) &           &               &              &                &          \\
\bottomrule
\end{tabular}
\label{tab:uip-weekly-borrows}
\end{table*}

\begin{table*}
\caption{Table of UIP results for weekly frequency data, using saving rates. 
Using Newey-West heteroscedasticity and autocorrelation robust standard errors (reported in parentheses.)}
\small
\begin{tabular}{lllllllcc}
        \toprule
Pair & N.obs & $\alpha$ & $\beta$ & R-squared & $\alpha$ p-value & $\beta$ p-value  &  Strict form (\ref{hyp:a}) p-value & Weak form (\ref{hyp:b}) p-value \\
\midrule
  eth\_bat &    56 &    0.04 &   -11.4299 &      0.21 &          0.01 &         0.00 &           0.00 &     0.00 \\
          &       &  (0.02) &     (1.51) &           &               &              &                &          \\
  eth\_zrx &    56 &    0.00 &   -2.21453 &      0.01 &          0.86 &         0.04 &           0.00 &     0.00 \\
          &       &  (0.02) &     (1.06) &           &               &              &                &          \\
 eth\_usdc &    56 &    0.01 &  -0.174619 &      0.00 &          0.84 &         0.83 &           0.01 &     0.16 \\
          &       &  (0.05) &     (0.83) &           &               &              &                &          \\
  eth\_dai &    25 &    0.08 &   -1.62795 &      0.10 &          0.25 &         0.35 &           0.28 &     0.14 \\
          &       &  (0.07) &     (1.73) &           &               &              &                &          \\
  eth\_sai &    56 &    0.00 &  -0.154556 &      0.01 &          0.79 &         0.65 &           0.00 &     0.00 \\
          &       &  (0.02) &     (0.34) &           &               &              &                &          \\
  eth\_rep &    56 &    0.01 &  -0.079075 &      0.00 &          0.48 &         0.84 &           0.03 &     0.01 \\
          &       &  (0.02) &     (0.40) &           &               &              &                &          \\
  bat\_zrx &    56 &   -0.01 &   -5.48743 &      0.07 &          0.41 &         0.21 &           0.20 &     0.15 \\
          &       &  (0.01) &     (4.39) &           &               &              &                &          \\
 bat\_usdc &    56 &    0.03 &  -0.841115 &      0.02 &          0.48 &         0.18 &           0.00 &     0.00 \\
          &       &  (0.04) &     (0.63) &           &               &              &                &          \\
  bat\_dai &    25 &    0.07 &   -1.73374 &      0.12 &          0.31 &         0.26 &           0.10 &     0.09 \\
          &       &  (0.07) &     (1.54) &           &               &              &                &          \\
  bat\_sai &    56 &    0.04 &  -0.817567 &      0.10 &          0.17 &         0.03 &           0.00 &     0.00 \\
          &       &  (0.03) &     (0.38) &           &               &              &                &          \\
  bat\_rep &    56 &    0.00 &    5.60551 &      0.14 &          0.77 &         0.00 &           0.00 &     0.00 \\
          &       &  (0.01) &     (0.39) &           &               &              &                &          \\
 zrx\_usdc &    56 &    0.04 &  -0.819943 &      0.01 &          0.53 &         0.46 &           0.00 &     0.10 \\
          &       &  (0.06) &     (1.10) &           &               &              &                &          \\
  zrx\_dai &    25 &    0.13 &   -2.35098 &      0.11 &          0.18 &         0.15 &           0.07 &     0.05 \\
          &       &  (0.10) &     (1.62) &           &               &              &                &          \\
  zrx\_sai &    56 &    0.04 &  -0.543762 &      0.03 &          0.08 &         0.09 &           0.00 &     0.00 \\
          &       &  (0.02) &     (0.32) &           &               &              &                &          \\
  zrx\_rep &    56 &    0.01 &    -1.1295 &      0.01 &          0.50 &         0.00 &           0.00 &     0.00 \\
          &       &  (0.02) &     (0.38) &           &               &              &                &          \\
 usdc\_dai &    25 &    0.00 &  -0.166452 &      0.18 &          0.10 &         0.10 &           0.00 &     0.00 \\
          &       &  (0.00) &     (0.10) &           &               &              &                &          \\
 usdc\_sai &    56 &    0.00 & -0.0475424 &      0.09 &          0.15 &         0.02 &           0.00 &     0.00 \\
          &       &  (0.00) &     (0.02) &           &               &              &                &          \\
 usdc\_rep &    56 &   -0.03 &  -0.866689 &      0.03 &          0.28 &         0.03 &           0.00 &     0.00 \\
          &       &  (0.03) &     (0.40) &           &               &              &                &          \\
  dai\_sai &    25 &   -0.02 &   -1.31716 &      0.07 &          0.22 &         0.36 &           0.24 &     0.12 \\
          &       &  (0.02) &     (1.43) &           &               &              &                &          \\
  dai\_rep &    25 &   -0.09 &   -1.72587 &      0.14 &          0.13 &         0.18 &           0.09 &     0.05 \\
          &       &  (0.06) &     (1.29) &           &               &              &                &          \\
\bottomrule
\end{tabular}
\label{tab:uip-weekly-saves}
\end{table*}

\clearpage
\twocolumn
\subsection{Vector Error Correction Models}
\label{sec:appendix-vecm}

Where time series are non-stationary (e.g.\ a random walk), the required criteria for a regression to produce be the Best Linear Unbiased Estimator (BLUE) are not satisfied, because the variables are not covariance stationary.\footnote{Covariance stationary means that the mean and autocovariances are finite and time invariant.}
However, if there exists a linear combination of non-stationary time series, where this combination is itself stationary, the series are said to be \textit{cointegrated}.
VECMs permit the modelling of the stationary relationships between such time series, and allow estimation of both the long-run and short-run adjustment dynamics. 
A VECM model is as follows. 

\begin{equation}
    \mathbf{\Delta y_t} = \mathbf{v + \Pi y_{t-1}} + \sum^{p-1}_{i=1} \mathbf{\Gamma_i \Delta y_{t-i}+ \epsilon_i}
\label{eqn:vecm}
\end{equation}

where $\Delta$ denotes a single time step, $\mathbf{y_t}$ is a vector of $K$ variables, $\mathbf{v}$ is a vector of $K \times 1$ parameters, $\mathbf{\Pi} = \sum^{j=p}_{j=1} \mathbf{A_j - I_k}$ ($I_k$ denotes an indicator vector), where $\mathbf{A_j}$ is a matrix of $K \times K$ parameters from a vector autoregression (VAR)\footnote{A VAR(p) can be expressed as $\mathbf{y_t = v + A_1y_{t-1} + A_2 y_{t-2} + ... + A_p y_{t-p} + \epsilon}$}, $\mathbf{\Gamma_i} = - \sum^{j=p}_{j=i+1} \mathbf{A_j}$ and $\epsilon$ is a $K \times 1$ vector of disturbances.
Assuming that $\Pi$ has reduced rank $0 < r< K$ it can further be expressed as $\mathbf{\Pi} = \mathbf{\alpha \beta'}$~\cite{statacorp2013stata}.
In terms of interpretation, $\mathbf{\alpha}$ provides the adjustment coefficients, $\mathbf{\beta}$ provides the parameters of the cointegrating (i.e.\ long-run) equations.

\clearpage

\begin{table}
\caption{Vector Error Correction Model Results - DAI.}
\label{tab:dai-vecm}

{
\def\sym#1{\ifmmode^{#1}\else\(^{#1}\)\fi}
\begin{tabular}{l*{1}{c}}
\toprule
                    &\multicolumn{1}{c}{(1)}\\
                    &\multicolumn{1}{c}{} \\
\midrule
D\_c\_dai             &                     \\
L.\_ce1              &     -0.0695         \\
                    &     (-1.20)         \\
\addlinespace
L.\_ce2              &       0.143         \\
                    &      (1.60)         \\
\midrule
D\_a\_dai             &                     \\
L.\_ce1              &       0.381\sym{***}\\
                    &      (4.66)         \\
\addlinespace
L.\_ce2              &      -0.533\sym{***}\\
                    &     (-4.23)         \\
\midrule
D\_d\_dai             &                     \\
L.\_ce1              &       0.284\sym{***}\\
                    &      (4.07)         \\
\addlinespace
L.\_ce2              &     -0.0387         \\
                    &     (-0.36)         \\

\midrule 
Long-run (\_ce1)               &                     \\
Compound DAI        &      1       \\
                &                \\
Aave DAI        &      0       \\
                &     (omitted)           \\
DyDx DAI       &    -1.151\sym{***}  \\
                 &     (-5.75) \\
Constant        &      0.0296       \\
                &              \\
Long-run (\_ce2)               &                     \\
Compound DAI        &      0       \\
                &          (omitted)      \\
Aave DAI        &      1      \\
                &               \\
DyDx DAI       &    -0.9906\sym{***}  \\
                        &     (-5.94) \\
Constant        &      0.0051       \\
                &              \\
\midrule
Observations        &         116         \\
\bottomrule
\multicolumn{2}{l}{\footnotesize \textit{t} statistics in parentheses}\\
\multicolumn{2}{l}{\footnotesize \sym{*} \(p<0.05\), \sym{**} \(p<0.01\), \sym{***} \(p<0.001\)}\\
\end{tabular}
}
\end{table}

\begin{figure}[H]
        \centering
        \includegraphics[width=0.47\textwidth]{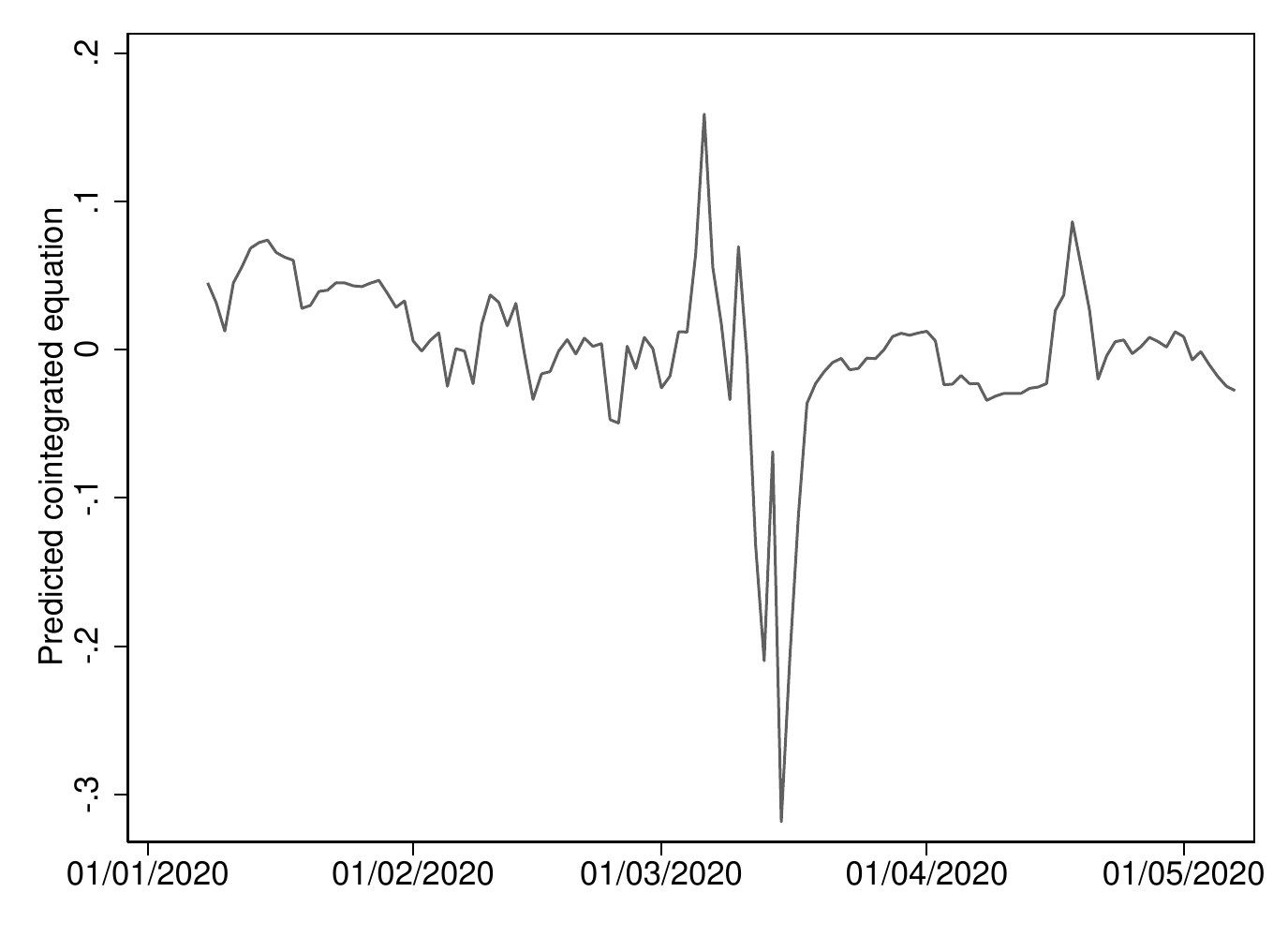}
        \caption{DAI cointegrating equation 1.}
        \label{fig:coint-dai1}
\end{figure}

\begin{figure}[H]
        \centering
        \includegraphics[width=0.47\textwidth]{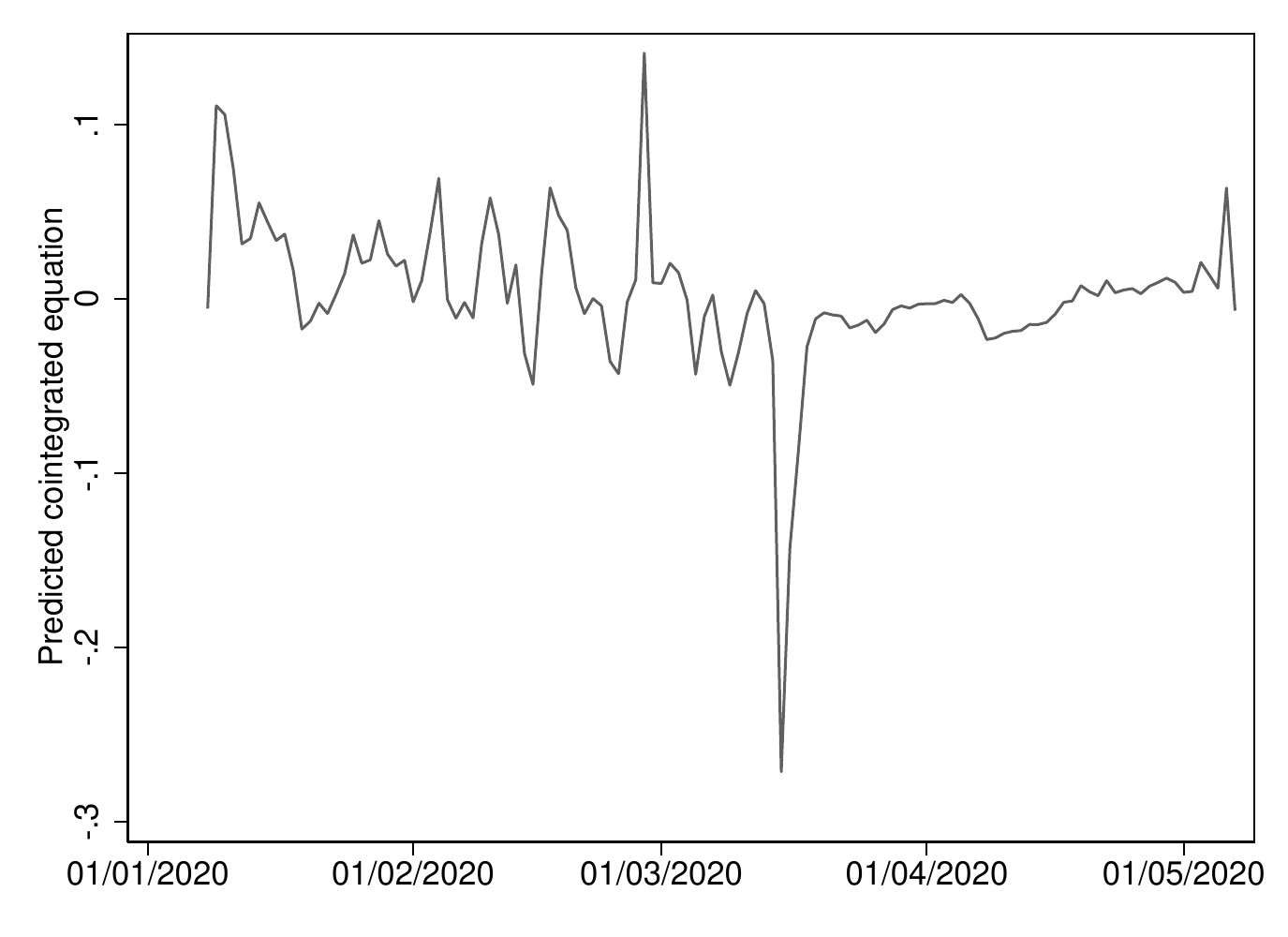}
        \caption{DAI cointegrating equation 2.}
        \label{fig:coint-dai2}
\end{figure}

\begin{figure}[H]
        \centering
        \includegraphics[width=\columnwidth]{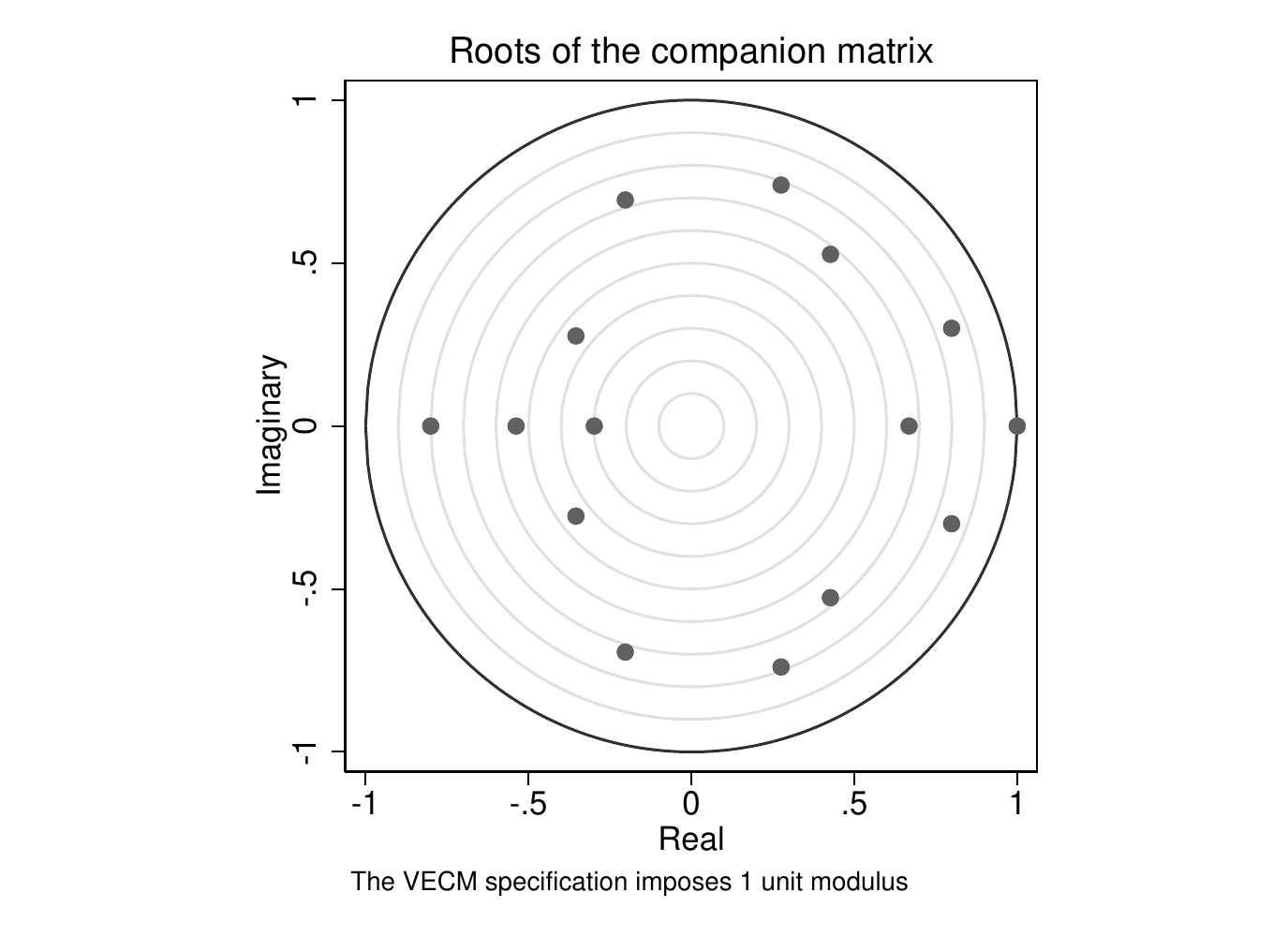}
        \caption{DAI cointegrating equations misspecification test.}
        \label{fig:dai-vecstable}
\end{figure}

\clearpage

\begin{table}
        \caption{Vector Error Correction Model Results - USDC.}
        \label{tab:usdc-vecm}
        
{
\def\sym#1{\ifmmode^{#1}\else\(^{#1}\)\fi}
\begin{tabular}{l*{1}{c}}
\toprule
                    &\multicolumn{1}{c}{(1)}\\
                    &\multicolumn{1}{c}{} \\
\midrule
D\_c\_usdc            &                     \\
L.\_ce1              &      0.0146         \\
                    &      (0.83)         \\
\addlinespace
L.\_ce2              &      0.0271         \\
                    &      (1.89)         \\
\midrule
D\_a\_usdc            &                     \\
L.\_ce1              &       0.607\sym{***}\\
                    &      (3.42)         \\
\addlinespace
L.\_ce2              &      -0.720\sym{***}\\
                    &     (-4.97)         \\
\midrule
D\_d\_usdc            &                     \\
L.\_ce1              &       0.115\sym{**} \\
                    &      (2.75)         \\
\addlinespace
L.\_ce2              &      0.0200         \\
                    &      (0.59)         \\
\midrule 
Long-run (\_ce1)               &                     \\
Compound USDC        &      1       \\
                &                \\
Aave USDC        &      5.55e-17       \\
                &     .           \\
DyDx USDC       &    -1.353\sym{***}  \\
                 &     (-7.77) \\
Constant        &      0.0066       \\
                &              \\
Long-run (\_ce2)               &                     \\
Compound DAI        &      -2.78e-17       \\
                &          .      \\
Aave DAI        &      1      \\
                &               \\
DyDx DAI       &    -1.347\sym{***}  \\
                        &     (-7.95) \\
Constant        &      0.00283       \\
                &              \\
\midrule
Observations        &         119         \\
\bottomrule
\multicolumn{2}{l}{\footnotesize \textit{t} statistics in parentheses}\\
\multicolumn{2}{l}{\footnotesize \sym{*} \(p<0.05\), \sym{**} \(p<0.01\), \sym{***} \(p<0.001\)}\\
\end{tabular}
}
\end{table}

\begin{figure}[H]
        \centering
        \includegraphics[width=0.47\textwidth]{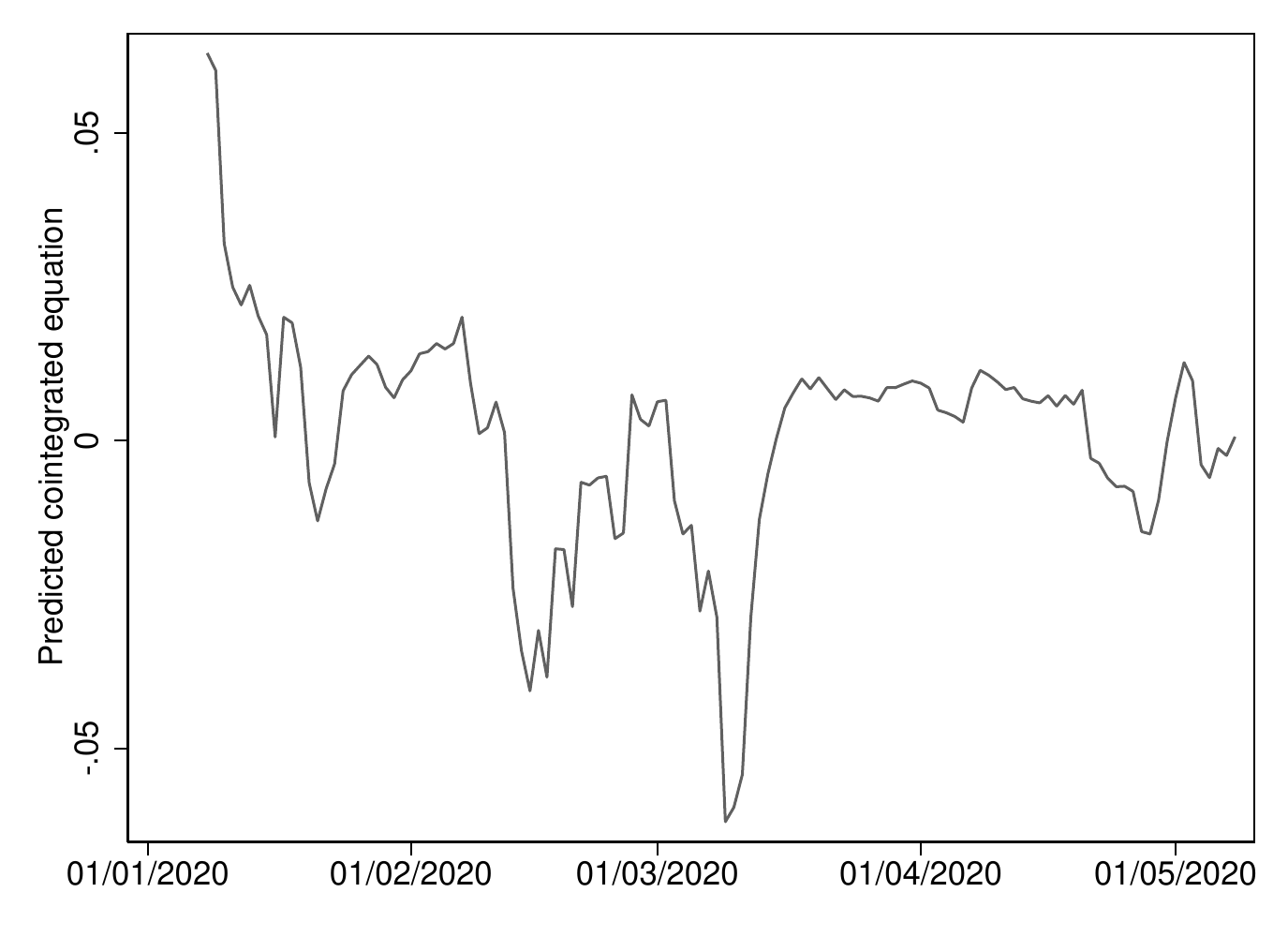}
        \caption{USDC cointegrating equation 1}
        \label{fig:coint-usdc1}
\end{figure}

\begin{figure}[H]
        \centering
        \includegraphics[width=0.47\textwidth]{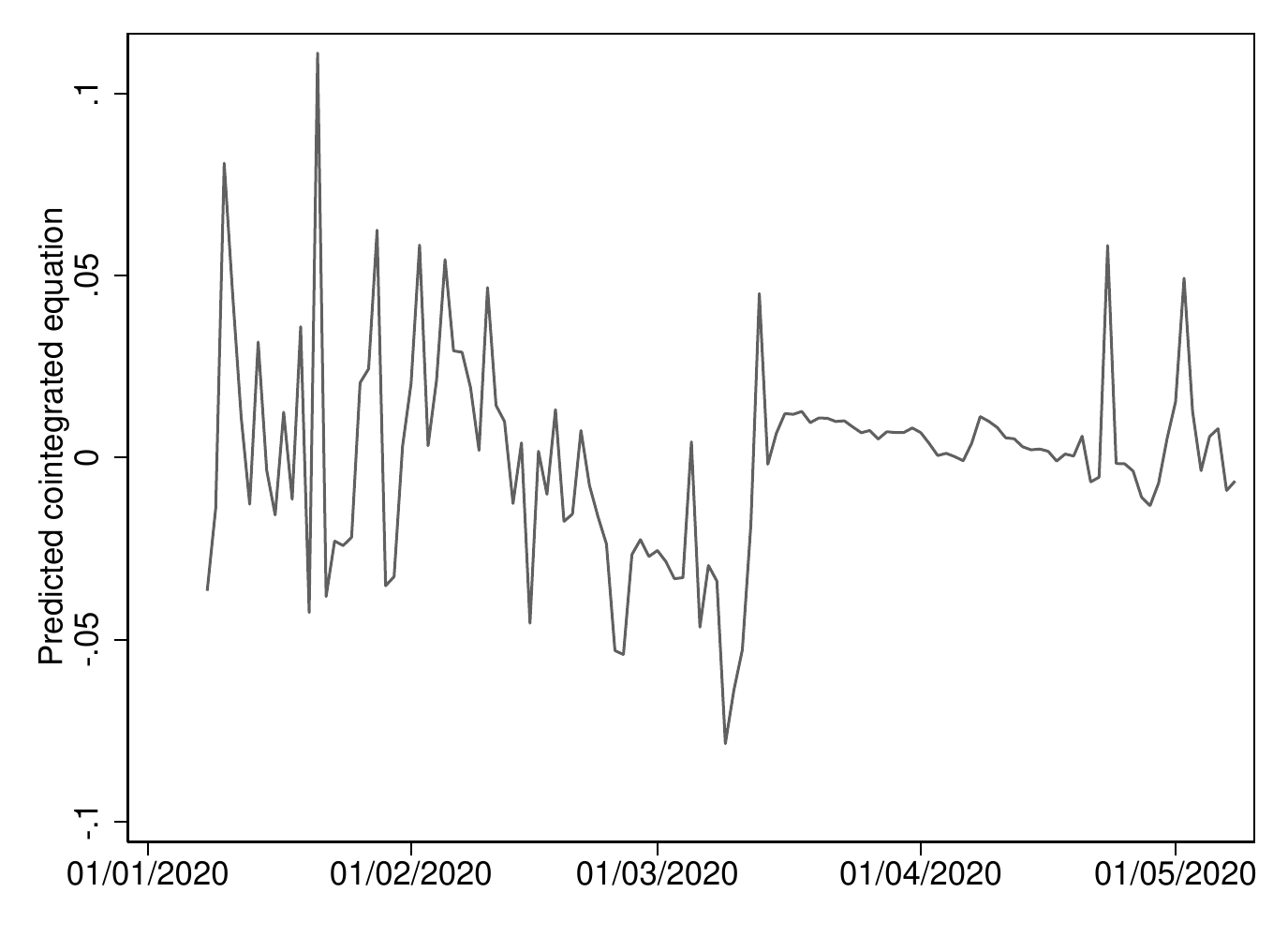}
        \caption{USDC cointegrating equation 2.}
        \label{fig:coint-usdc2}
\end{figure}

\begin{figure}[H]
        \centering
        \includegraphics[width=\columnwidth]{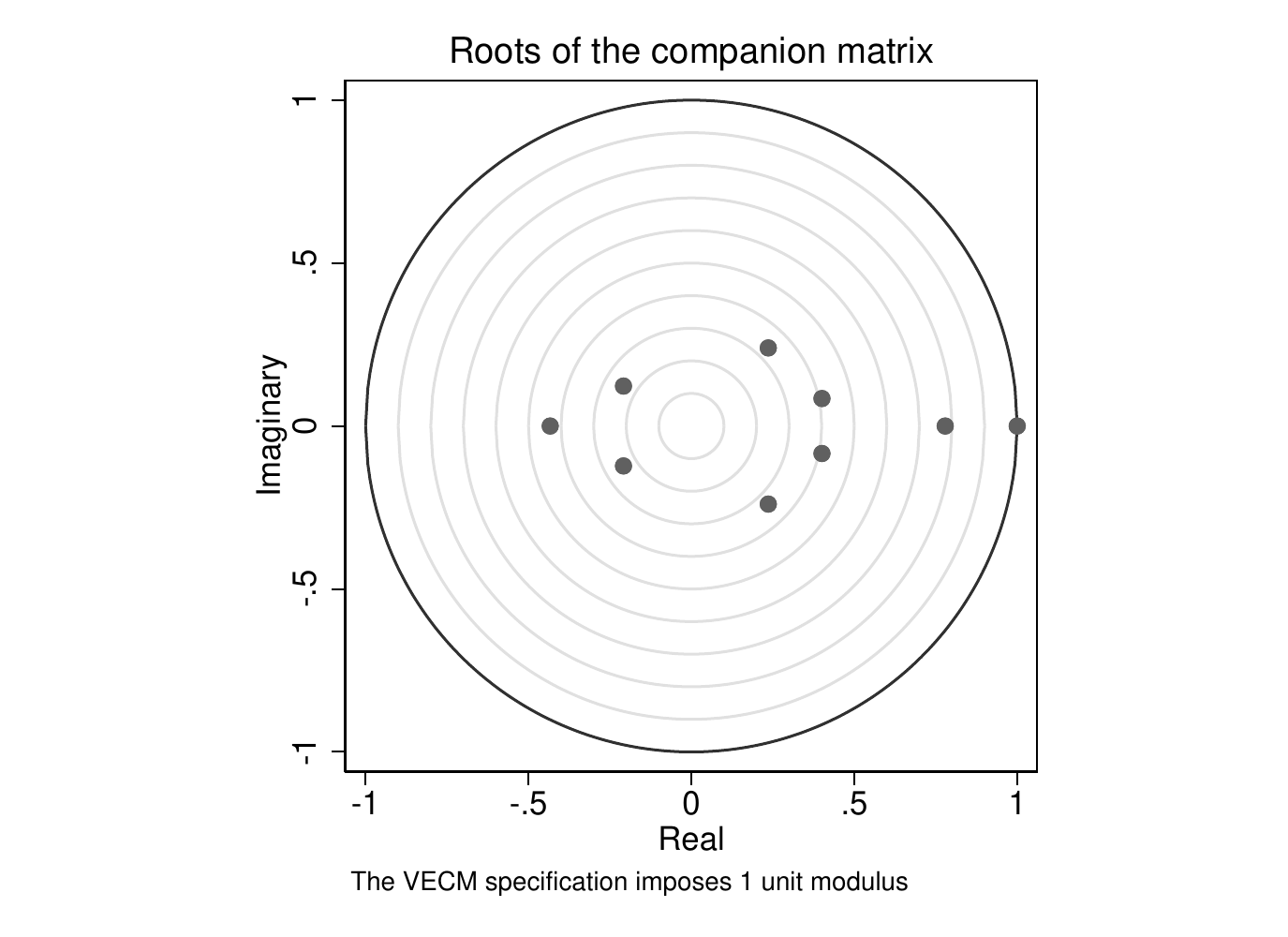}
        \caption{USDC cointegrating equations misspecification test.}
        \label{fig:usdc-vecstable}
\end{figure}

\end{document}